\documentclass[a4paper,11pt]{article}
\pdfoutput=1
\usepackage{jheppub}
\usepackage{slashed}
\usepackage[dvipsnames,table]{xcolor}
\usepackage{hyperref} 
\usepackage{xspace}
\usepackage[tight]{subfigure}
\usepackage{amsmath}
\usepackage{amssymb}
\usepackage{amsfonts}
\usepackage{mathrsfs}
\usepackage{comment}
\usepackage{afterpage}
\usepackage{verbatim}
\usepackage{booktabs}
\usepackage{array}
\usepackage[title,titletoc]{appendix}
\usepackage{tabularx}
\newcolumntype{C}[1]{>{\centering\arraybackslash}p{#1}}

\def\refeq#1{\mbox{(\ref{#1})}}

\def\citeres#1{\mbox{Refs.~\cite{#1}}}

\newcommand{\newc}{\newcommand}
\newc{\beq}{\begin{equation}}
\newc{\eeq}{\end{equation}}
\newc{\bit}{\begin{itemize}}
\newc{\eit}{\end{itemize}}
\newc{\ben}{\begin{enumerate}}
\newc{\een}{\end{enumerate}}
\newc{\bce}{\begin{center}}
\newc{\ece}{\end{center}}
\newc{\bfi}{\begin{figure}}
\newc{\efi}{\end{figure}}

\newcommand{\ri}{\mathrm i}
\newcommand{\re}{\mathrm e}
\newcommand{\rd}{\mathrm d}

\newcommand{\rT}{{\mathrm{T}}}
\newcommand{\rR}{{\mathrm{R}}}
\newcommand{\rL}{{\mathrm{L}}}
\newcommand{\rF}{{\mathrm{F}}}

\newcommand{\ie}{\emph{i.e.}\ }
\newcommand{\eg}{\emph{e.g.}\ }

\newcommand{\GeV}{\ensuremath{\,\text{GeV}}\xspace}
\newcommand{\TeV}{\ensuremath{\,\text{TeV}}\xspace}

\newcommand{\qqb}{{q\bar{q}}\xspace}
\newcommand{\PH}{\ensuremath{\text{H}}\xspace}
\newcommand{\Pj}{\ensuremath{\text{j}}\xspace}
\newcommand{\Pp}{\ensuremath{\text{p}}}
\newcommand{\Pe}{\ensuremath{\text{e}}\xspace}
\newcommand{\Pb}{\ensuremath{\text{b}}\xspace}

\newcommand{\Pt}{\ensuremath{\text{t}}\xspace}
\newcommand{\Pu}{\ensuremath{\text{u}}\xspace}
\newcommand{\Pd}{\ensuremath{\text{d}}\xspace}
\newcommand{\Ps}{\ensuremath{\text{s}}\xspace}
\newcommand{\Pc}{\ensuremath{\text{c}}\xspace}
\newcommand{\Pg}{\ensuremath{\text{g}}}

\newcommand{\PW}{\ensuremath{\text{W}}\xspace}
\newcommand{\PZ}{\ensuremath{\text{Z}}\xspace}

\newcommand{\Mt}{\ensuremath{m_\Pt}\xspace}
\newcommand{\MH}{\ensuremath{M_\PH}\xspace}

\newcommand{\MW}{\ensuremath{M_\PW}\xspace}

\newcommand{\Gt}{\ensuremath{\Gamma_\Pt}\xspace}
\newcommand{\GH}{\ensuremath{\Gamma_\PH}\xspace}

\newcommand{\GF}{\ensuremath{G_\mu}}

\newcommand{\MVOS}{\ensuremath{M_{V}^\text{OS}}\xspace}%
\newcommand{\GVOS}{\ensuremath{\Gamma_{V}^\text{OS}}\xspace}%

\newcommand{\recola}{{\sc Recola}\xspace}

\newcommand{\mocanlo}{{\sc MoCaNLO}\xspace}

\newcommand{\madgraph}{{\sc\small MadGraph}\xspace}

\newcommand{\madspin}{{\sc\small MadSpin}\xspace}

\newcommand{\phantommc}{{\sc\small PHANTOM}\xspace}

\newcolumntype{.}{D{.}{.}{-1}}
\newcolumntype{d}[1]{D{.}{.}{#1}}
\colorlet{tableoverheadcolor}{gray!37.5}
\colorlet{tableheadcolor}{gray!25}
\colorlet{tablerowcolor}{gray!12.5}

\def\draftdate{\relax}
\def\mda{\relax}
\def\mua{\relax}
\def\mla{\relax}
\def\draft{
\def\thtystars{******************************}
\def\sixtystars{\thtystars\thtystars}
\typeout{}
\typeout{\sixtystars**}
\typeout{* Draft mode!
         For final version remove \protect\draft\space in source file *}
\typeout{\sixtystars**}
\typeout{}
\def\draftdate{\today}
\def\mua{\marginpar[\boldmath\hfil$\uparrow$]%
                   {\boldmath$\uparrow$\hfil}\color{black}%
                    \typeout{marginpar: $\uparrow$}\ignorespaces}
\def\mda{\color{red}\marginpar[\boldmath\hfil$\downarrow$]%
                   {\boldmath$\downarrow$\hfil}%
                    \typeout{marginpar: $\downarrow$}\ignorespaces}
\def\mla{\marginpar[\boldmath\hfil$\rightarrow$]%
                   {\boldmath$\leftarrow $\hfil}%
                    \typeout{marginpar: $\leftrightarrow$}\ignorespaces}
\def\Mua{\marginpar[\boldmath\hfil$\Uparrow$]%
                   {\boldmath$\Uparrow$\hfil}\color{black}%
                    \typeout{marginpar: $\uparrow$}\ignorespaces}
\def\Mda{\color{red}\marginpar[\boldmath\hfil$\Downarrow$]%
                   {\boldmath$\Downarrow$\hfil}%
                    \typeout{marginpar: $\downarrow$}\ignorespaces}
\def\Mla{\marginpar[\boldmath\hfil\textcolor{red}{$\Rightarrow$}]%
                   {\boldmath\textcolor{red}{$\Leftarrow $}\hfil}%
                    \typeout{marginpar: $\leftrightarrow$}\ignorespaces}
\overfullrule 5pt
\oddsidemargin 15mm
\marginparwidth 29mm
}

\newcommand{\eqn}[1]{Eq.~(\ref{#1})}

\newcommand{\tab}[1]{Table~\ref{#1}}

\newcommand{\fig}[1]{Fig.~\ref{#1}}

\newcommand{\figsa}[2]{Figs.~\ref{#1} and \ref{#2}}
\newcommand{\sect}[1]{Sect.~\ref{#1}}

\newcommand{\rf}[1]{Ref.~\cite{#1}}

\newcommand{\rfs}[1]{Refs.~\cite{#1}}  

\newcommand{\mc}{\mathcal}

\newcommand{\Mw}{M_{\rm{W}}}
\newcommand{\Mz}{M_{\rm{Z}}}

\newcommand{\Gw}{\Gamma_{\rm{W}}}

\newcommand{\Mwo}{M^{\rm os}_{\rm{W}}}
\newcommand{\Mzo}{M^{\rm os}_{\rm{Z}}}
\newcommand{\Gzo}{\Gamma^{\rm os}_{\rm{Z}}}
\newcommand{\Gwo}{\Gamma^{\rm os}_{\rm{W}}}

\newcommand{\as}{\alpha_{\textrm{s}}}
\newcommand{\pt}{p_{\rT}}
\newcommand{\ptj}{p_{\rT,\Pj}}
\newcommand{\ptb}{p_{\rT,\Pb}}
\newcommand{\ptmiss}{p_{\rT,{\rm miss}}}
\newcommand{\pte}{p_{\rT,\Pe^+}}
\newcommand{\ptl}{p_{\rT,\ell}}

\newcommand{\enmn}{\Pe^+\nu_\Pe\mu^-\bar\nu_\mu}
\newcommand{\ppenmn}{\Pp\Pp\to\enmn+X}

\newcommand{\ggenmn}{\Pg\Pg\to\enmn}

\title{Polarized electroweak bosons in ${\bf \PW^+\PW^-}$ production at the LHC including NLO QCD effects} 
\author{Ansgar Denner and}
\author{Giovanni Pelliccioli}
\affiliation{Universit\"at W\"urzburg, Institut f\"ur Theoretische Physik und Astrophysik, 97074 W\"urzburg, Germany}
\emailAdd{ansgar.denner@physik.uni-wuerzburg.de}
\emailAdd{giovanni.pelliccioli@physik.uni-wuerzburg.de}
\date{\draftdate}

\abstract{The measurement of polarization fractions of massive gauge
  bosons at the LHC provides an important check of the Standard Model
  and in particular of the Electroweak Symmetry Breaking mechanism.
  Owing to the unstable character of $\PW$ and $\PZ$~bosons, devising
  a theoretical definition for polarized signals is not
  straightforward and always subject to some ambiguity.  Focusing on
  $\PW$-boson pair production at the LHC in the fully leptonic
  channel, we propose to compute polarized cross-sections and
  distributions based on the gauge-invariant doubly-resonant part of
  the amplitude.  We include NLO QCD corrections to the leading
  quark-induced partonic process and also consider the loop-induced
  gluon-initiated process contributing to the same final state.  We
  present results for both an inclusive setup and a realistic fiducial
  region, with special focus on variables that are suited for the
  discrimination of polarized cross-sections and on quantities that
  can be measured experimentally.  }


\begin{document}

\strut\hfill\draftdate

\maketitle
\section{Introduction}

The production of electroweak (EW) gauge bosons at the Large Hadron
Collider (LHC) has been investigated intensively in recent years both
theoretically and experimentally.

During Run 2 at $13\TeV$ centre-of-mass (CM) energy, the LHC
accumulated enough luminosity to enable precise measurements of a
large variety of multi-boson processes both in the light of probing
the Standard Model (SM) of fundamental interactions and with the aim
of hunting for new physics.  The accumulated statistics allows for the
measurement of observables that are difficult to extract from the
data, like the polarizations of massive gauge bosons.  Being unstable,
$\PW$ and $\PZ$~bosons can only be produced off-shell and
reconstructed from their hadronic or leptonic decay products.
Therefore, the only way to get access to their polarization is to
study their decay products, with a particular focus on the angular
variables that have a direct dependence on the polarization mode of
the decayed boson.

The analysis of polarization could serve as a probe of the SM
gauge and Higgs sectors as well as a tool to discriminate between
the SM and beyond-the-Standard-Model (BSM) theories.  In particular,
since the longitudinal polarization is a direct consequence of the
Electroweak Symmetry Breaking (EWSB) mechanism, any deviation from the
SM in the production of longitudinal weak bosons would provide
valuable information.

A number of polarization measurements has been performed with LHC data
at $8\TeV$ CM energy.  The CMS and ATLAS collaborations have extracted
polarization fractions and coefficients of angular distributions in
$\PW$-boson production in association with jets
\cite{Chatrchyan:2011ig,ATLAS:2012au}, in inclusive $\PZ$-boson
production \cite{Aad:2016izn,Khachatryan:2015paa}, and in
$\Pt\bar{\Pt}$ events \cite{Aaboud:2016hsq, Khachatryan:2016fky}.  A
combination of CMS and ATLAS $\PW$-boson polarization measurements in
top-quark decays appeared very recently \cite{Aad:2020jvx}.  The first
polarization measurement with $13\TeV$ data has been performed by
ATLAS in $\PW^\pm\PZ$~production \cite{Aaboud:2019gxl}.  The increased
luminosities expected in the high-luminosity run of the LHC will
enable polarization measurements even in processes with rather small
cross-sections, like vector-boson scattering \cite{CMS:2018mbt,
  Azzi:2019yne}.

Several theoretical results on polarized weak bosons at the LHC are
available in the literature.  A detailed study of $\PW$-boson
polarization in $\PW+{}$jet production has been performed in
\rf{Bern:2011ie} in the absence of lepton cuts. The effect of
realistic selection cuts has been studied in \rf{Stirling:2012zt} both
in $V+{}$jets and in many other multi-boson production processes,
including a few results at leading-order (LO) for $\PW\PW$ and
$\PW\PZ$~production.  The effect of selection cuts and their interplay
with interferences between amplitudes for different polarizations has
been investigated in \rf{Belyaev:2013nla}. The polarization of $\PW$
and $\PZ$~bosons in vector-boson scattering has been extensively
studied in the fully leptonic channel at LO EW in
\rfs{Ballestrero:2017bxn,Ballestrero:2019qoy} using the \phantommc
Monte Carlo \cite{Ballestrero:2007xq}.  Recently, fiducial
polarization observables have been analysed in fully leptonic
$\PW^\pm\PZ$ production including next-to-leading (NLO) QCD and EW
corrections both in a realistic LHC environment \cite{Baglio:2018rcu}
and in an inclusive setup \cite{Baglio:2019nmc}.  In particular,
angular coefficients that can be extracted analytically from
unpolarized distributions are directly related to the polarization
fractions in the absence of lepton cuts.
  Extending this strategy in the presence of lepton cuts provides
  fiducial observables that, while being accessible at the LHC, can
  be very far from describing the polarization of decayed weak bosons.
The calculation of polarized
cross-sections has been automated in the \madgraph Monte Carlo
\cite{BuarqueFranzosi:2019boy}, employing decay chains in the
narrow-width approximation and including spin correlations via the
\madspin package \cite{Artoisenet:2012st}.  Very recently, the
gluon-induced $\PZ\PZ$ production with polarized bosons has been
studied with the aim of enhancing the sensitivity to the
$\PZ\Pt\bar{\Pt}$ coupling \cite{Cao:2020npb}.

In this paper, focusing the discussion on the $\PW^+\PW^-$ production
at the LHC@13TeV, we propose a method to define signals with polarized
weak bosons at the amplitude level, including also QCD radiative
corrections. This actually represents an extension to NLO QCD of the
method developed in vector-boson scattering with the \phantommc Monte
Carlo \cite{Ballestrero:2017bxn,Ballestrero:2019qoy}.
The definition of polarized signals presented in this work relies on
the double-pole approximation (DPA)
\cite{Aeppli:1993cb,Aeppli:1993rs,Beenakker:1998gr,Denner:2000bj,%
Billoni:2013aba,Biedermann:2016guo},
which is expected to be more accurate than the narrow-width
approximation or a decay chain.

The production of $\PW^+\PW^-$ pairs in the fully leptonic decay
channel has been extensively studied at the theoretical level.  Beyond
its own importance as a clean signature at hadron colliders, it is an
important irreducible background for Higgs searches, and provides a
handle to probe the SM triple gauge-boson coupling. It has also been
widely investigated in the context of direct BSM searches.  The SM
radiative corrections to the full process with leptonic decays are
known up to next-to-next-to leading-order (NNLO) QCD and NLO EW
\cite{Bierweiler:2012kw,
  Caola:2015rqy,Grazzini:2016ctr,Biedermann:2016guo,Kallweit:2019zez}.
All order resummation and parton-shower effects have also been studied
\cite{Re:2018vac,Kallweit:2020gva,Brauer:2020kfv,Chiesa:2020ttl,Kuhn:2011mh}.
NLO QCD results on polarized $\PW^+\PW^-$ production within the
SM and the Standard Model Effective Field Theory are shown in \rf{Baglio:2017bfe}
for on-shell bosons \PW~bosons.

However, a detailed phenomenological analysis of polarized \PW-boson
pair production including off-shell effects 
is still missing in literature.  The computation we
present includes NLO QCD corrections to the leading $\qqb$ partonic
channel, and the LO predictions for the loop-induced $\Pg\Pg$ partonic
channel, which for the first time is studied in its polarization
structure.  Providing accurate SM predictions for polarized di-boson
production is of great importance both to further probe the SM itself
and to investigate possible deviations from the SM triple gauge-boson
couplings and other potential new-physics effects.

There is a long-standing experimental interest in the polarizations of
$\PW$~bosons in boson-pair production. They were investigated for
example in electron--positron collisions at LEP \cite{opal1999} with
the aim of probing anomalous gauge-boson couplings.

At the LHC, $\PW$-pair production has been measured by ATLAS
\cite{Aaboud:2017qkn,Aaboud:2019nkz} with $13\TeV$ data, without the
specific aim of extracting polarization fractions. Differently from
$\PW^\pm \PZ$ and $\PZ\PZ$~production, the presence of two neutrinos
in the final state hampers the reconstruction of the two vector bosons,
reducing the number of variables that discriminate among
polarization states. However, the accumulated luminosity in Run 2
allows for the extraction of polarizations, \eg by means of a
multi-variate analysis of relevant observables based on SM Monte
Carlo templates. Furthermore, the recent ATLAS results in $\PW^\pm
\PZ$ production \cite{Aaboud:2019gxl} and the foreseen measurements of
polarizations in $\PW\PW$~scattering with high-luminosity data
\cite{Azzi:2019yne} give us confidence that polarizations will soon be
investigated in $\PW$-pair production.

This paper is organized as follows. In \sect{sub:def} we describe the
strategy we use to define polarized cross-sections in $\PW$-pair
production. In \sect{sub:set} we present the setup of our simulations,
including SM parameters and selection cuts. Results for the production
of one or two polarized bosons in an inclusive setup are shown in
\sect{sub:inc}, while corresponding results in a realistic
fiducial region are compiled in \sect{sub:fid}. We present both total
cross-sections and differential distributions with a focus on the
impact of QCD corrections on polarization fractions and, more in
general, on polarized distributions for the relevant kinematic
variables and observables. In \sect{sub:con} we draw our conclusions.

\section{Details of the calculation}
\subsection{Definition of the polarized signals}\label{sub:def}

A precise definition of polarized signals can be given only for
on-shell particles.  Since $\PW$ and $\PZ$~bosons are unstable
particles, selecting their polarization states is always afflicted
with some ambiguity.
In this work, we define polarized cross-sections via an extension of
the techniques used at LO in
\rfs{Ballestrero:2017bxn,Ballestrero:2019qoy} to NLO QCD.  It is worth
recalling the main steps taken to arrive at a well-behaved definition of
polarized EW bosons at the amplitude level.

In all the following we focus on $\PW$-pair production, but everything
is valid without modifications for any boson-pair production process
in the fully leptonic decay channel, given that the two pairs of
leptons have different flavours. Considering same flavour lepton pairs
would require a more or less trivial extension of the DPA. The
strategy can be directly transferred to define polarized cross
sections for the production of arbitrary unstable particles.

\begin{figure}
  \centering
  \subfigure[Resonant diagram\label{subfig:res}]{\hspace{0.8cm}\includegraphics[scale=0.7]{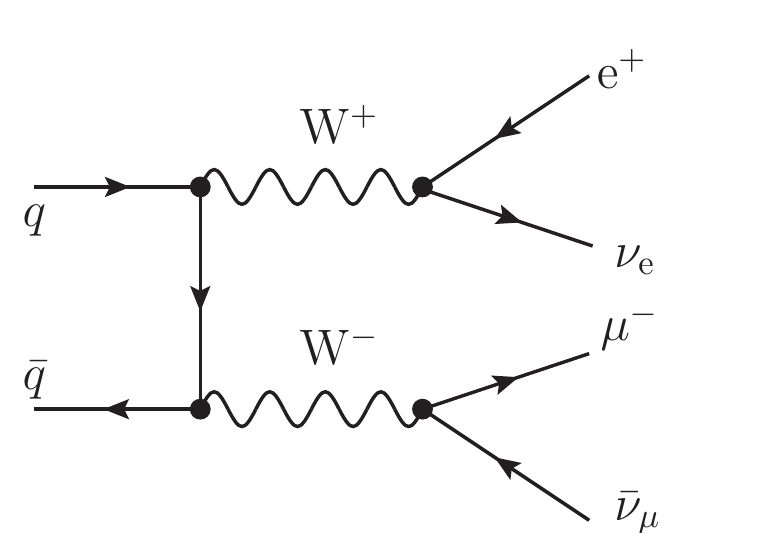}}
  \subfigure[Non-resonant diagram\label{subfig:nonres}]{\hspace{1.2cm}\includegraphics[scale=0.7]{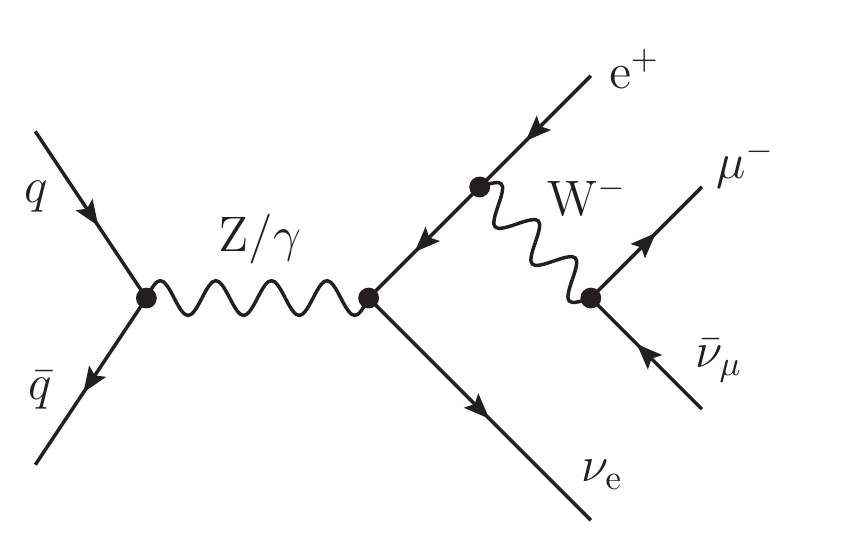}}
  \caption{Sample tree-level diagrams for $\PW^+\PW^-$ production at the LHC.}\label{fig:diagww}
\end{figure}

The first obstacle towards the definition of polarized signals are
 the non-resonant contributions to the full process $\ppenmn$.  Already
at tree level in the SM (LO EW), di-boson production receives
contributions from both resonant [\fig{subfig:res}] and
non-resonant diagrams [\fig{subfig:nonres}].
The former diagrams give rise to an amplitude that can be factorized
into production, propagation, and decay of vector bosons, \ie they
contain a contribution to vector-boson pair production. The latter
diagrams do not involve two intermediate vector bosons, and a
definition of corresponding polarizations does not make sense. In
fact, non-doubly-resonant diagrams cannot be viewed as a contribution
to vector-boson pair production, but should be treated as irreducible 
background and subtracted from the complete process.  However, since
non-resonant contributions are necessary to preserve gauge invariance,
they cannot be dropped without further manipulating the resulting
amplitude.

This issue is easily solved by using a narrow-width approximation,
which however can be very inaccurate, since off-shell effects and spin
correlations are completely neglected. Beyond simple decay-chain
techniques, a better solution is provided by the
\madspin method
\cite{Artoisenet:2012st,BuarqueFranzosi:2019boy} that preserves LO
spin correlations and reintroduces an off-shell-ness of weak-boson
propagators.  Another approach, which has proven to be accurate in
vector-boson scattering and other multi-boson signatures, is given by
the DPA
\cite{Aeppli:1993cb,Aeppli:1993rs,Beenakker:1998gr,Denner:2000bj,%
Billoni:2013aba,Biedermann:2016guo}
or pole approximations in general \cite{Stuart:1991xk,Denner:2019vbn}:
given a resonant amplitude, the numerator of the resonant diagrams is
projected on shell to restore gauge invariance, while the Breit--Wigner
modulation is kept with off-shell kinematics. This procedure can be
viewed as a definition of the di-boson production contribution and
provides a gauge-invariant separation of the complete process into 
$\PW$-pair production and the corresponding irreducible background.
As the pole approximation, this separation does not only work for
$\PW$-pair production but for arbitrary processes involving unstable
particles. This technique has been employed to obtain the results
presented in this paper.

Once $\PW$-pair production is properly isolated via the DPA in a
gauge-invariant manner, individual polarized contributions can be
defined. Note, however, that polarizations of massive states are not
uniquely fixed.  Let us consider a generic amplitude that involves the
production of a $\PW$~boson with momentum $k$ (subamplitude $\mc
P_{\mu}$) and its decay into massless leptons with momenta $l$ and
$k-l$ (subamplitude $\mc D_{\nu}$). In the 't~Hooft--Feynman gauge
such an amplitude reads
\beq
\mc A_{\rm res} = \mc P_{\mu}(k)\,\frac{-g^{\mu\nu}}{k^2-\Mw^2+\ri\Gw\Mw}\,\mc D_{\nu}(l,k-l)\,.
\eeq
Using polarization vectors $\varepsilon_\lambda^{\mu}$ for the
$\PW$~boson, the on-shell propagator numerator can be rewritten as
follows,
\begin{align}\label{eq:seppol}
\mc A_{\rm res}
={}&
\mc
P_{\mu}(k)\,\frac{\big[\sum_{\lambda=\rL,\pm}\varepsilon_\lambda^{\mu\,*}(k)\varepsilon_\lambda^{\nu}(k)\big]\,-k^\mu k^\nu/\Mw^2\,}{k^2-\Mw^2+\ri\Gw\Mw}\,\mc
D_{\nu}(l,k-l)
\notag\\={}&\sum_{\lambda=\rL,\pm}\,\frac{\mc
  P_{\mu}(k)\varepsilon_\lambda^{\mu\,*}(k)\,\,\,\varepsilon_\lambda^{\nu}(k)\mc D_{\nu}(l,k-l)}{k^2-\Mw^2+\ri\Gw\Mw}\,-\,
\frac{\mc
  P_{\mu}(k) k^{\mu}\,\,\,k^{\nu}\mc D_{\nu}(l,k-l)}{\Mw^2(k^2-\Mw^2+\ri\Gw\Mw)}
\notag\\={}&
\sum_{\lambda=\rL,\pm}\,\frac{\mc M_{\lambda}(k)\,\mc M_{\lambda}(l,k-l)}{k^2-\Mw^2+\ri\Gw\Mw}\,
+ 0
=:\, \sum_{\lambda=\rL,\pm} \mc A_\lambda\,,
\end{align}
where the sum runs over the three physical polarizations, longitudinal
($\lambda=\rL$), left-handed ($\lambda=-$), and right-handed
($\lambda=+$).  The term proportional to the boson momentum
$k$ vanishes upon contraction with the massless leptonic current $\mc
D_{\nu}$ [second line of \eqn{eq:seppol}].  Then, the numerator can be
expressed as a sum over products of polarized matrix elements for the
production [$\mc M_{\lambda}(k)$] and the decay [$\mc
M_{\lambda}(l,k-l)$] of the polarized \PW~bosons.  Note also that
would-be Goldstone-boson diagrams vanish thanks to the massless decay
leptons.  These simplifications hold within any $R_\xi$ gauge. We
stress that if the final leptons are massive, the additional term
proportional to the boson momentum in \eqn{eq:seppol} cancels against
the would-be Goldstone-boson contributions, leaving 
the same sum over the physical polarization states.

It is worth noting that the polarization vectors introduced in
\eqn{eq:seppol} need to be defined in a specific reference frame. In
all results presented in this paper, we choose the laboratory frame
for this purpose. To be precise, for an on-shell vector boson with mass~$M$,
energy $E$, and momentum $p=\sqrt{E^2-M^2}$ that propagates along the
direction defined by the spherical angles $\theta_V$ and $\phi_V$, the
polarizations vectors read
\begin{align}
\varepsilon^{\mu}_{-} &= \frac{1}{\sqrt 2}
(0,\cos\theta_V\cos\phi_V +  \ri\sin\phi_V , \cos\theta_V\sin\phi_V - \ri\cos\phi_V, - \sin\theta_V) \,,\notag \\
\varepsilon^{\mu}_{+} &= \frac{1}{\sqrt 2}
(0,- \cos\theta_V\cos\phi_V +  \ri\sin\phi_V , - \cos\theta_V\sin\phi_V - \ri\cos\phi_V, \sin\theta_V) \,,\notag\\
\varepsilon^{\mu}_{\rL} &=
\frac{1}{M}(p,E \sin\theta_V\cos\phi_V,E \sin\theta_V\sin\phi,E \cos\theta_V)\,.
\end{align}


To obtain polarized cross-sections we need to square \eqn{eq:seppol}.
The result is not simply the sum of squared polarized terms, since
interferences among different polarization states arise,
\beq\label{eq:interfdef}
\big|\mc A_{\rm res}\big|^2 = \sum_\lambda \big|\mc A_\lambda\big|^2 + \sum_{\lambda\neq \lambda'}\mc A_\lambda^*\,A_{\lambda'}\,.
\eeq
Such interferences [the second term in \eqn{eq:interfdef}] are
expected to vanish only for fully inclusive decays, \ie in the absence
of cuts on the decay leptons, but are non-zero otherwise
\cite{Stirling:2012zt,Belyaev:2013nla,Ballestrero:2017bxn}.  However,
interference effects play a non-negligible role even in an inclusive
setup for \PW-pair production.  From \eqn{eq:interfdef} it is evident
that with a simple substitution in the propagator,
\beq
\sum_{\lambda'}\varepsilon_{\lambda'}^{\mu\,*}\varepsilon_{\lambda'}^{\nu} \,\longrightarrow\, \varepsilon_\lambda^{\mu\,*}\varepsilon_\lambda^{\nu}\,,
\eeq
for a given polarization state $\lambda$ ($={\rL},-,+$), we are able to
compute polarized cross-sections with a Monte Carlo.  The size of the
interferences can then be easily deduced by comparing the unpolarized
results [l.h.s.\ of \eqn{eq:interfdef}] with the sum of the polarized
ones [first term on the r.h.s\ in \eqn{eq:interfdef}].

Since in $\PW^+\PW^-$ production there are two vector bosons, not only
singly-polarized but also doubly-polarized configurations can be
investigated.  To compute the doubly-polarized signals, we select a
definite polarization state for both bosons, resulting in 9 different
possible combinations. However, as usually done in experimental
analyses, the left- and right-handed contributions of a single vector
boson are combined by means of a coherent sum into the transverse one,
which also includes the left--right interference term (which is
non-zero in general).  In the end we are left with four different
combinations of longitudinal (L) and transverse (T) modes: 
$\PW^+_{\rL}\PW^-_{\rL}$, $\PW^+_{\rL}\PW^-_{\rT}$,
$\PW^+_{\rT}\PW^-_{\rL}$, and $\PW^+_{\rT}\PW^-_{\rT}$.  When computing
singly-polarized signals, one boson has fixed polarization state (L or T), while
the other one is kept unpolarized, \ie we include the coherent sum of
all of its polarization modes. In the following we consider
singly-polarized results for the $\PW^+$ boson.

Finally, we include NLO QCD corrections.  As detailed above, 
we use the DPA to define polarized cross-sections.  In the
literature, the DPA is usually applied only to contributions that
feature Born-level kinematics, such as the LO or virtual corrections.
In order to define polarized \PW~bosons in all NLO contributions, we
have to apply the DPA also to the real QCD corrections.
Since we employ the Catani--Seymour scheme for the subtraction of
infrared singularities \cite{Catani:1996vz}, it has to be applied
also to integrated and unintegrated subtraction dipoles, which turned
out to be the most delicate part of the calculation. 

The extension of this strategy to NLO EW is possible though slightly
more complicated. Since both virtual and real corrections lead to
non-factorizable but resonant contributions and real photons can be
radiated off the resonant boson, various contributions have to be
taken into account in the DPA. For each doubly-resonant term,
polarized amplitudes have to be defined, and each term that is
non-doubly-resonant has to be split off as background. This can be
done following the line of reasoning in
\citeres{Beenakker:1997ir,Denner:1997ia,Beenakker:1998gr}.

QCD radiative corrections only modify the production
subprocess in di-boson production, leading to an easier implementation
in a numerical code.

All results presented in this work have been obtained using \recola
amplitudes \cite{Actis:2012qn, Actis:2016mpe} and \mocanlo, which is a
multi-channel Monte Carlo integration code that has already been used
for several calculations at NLO QCD and EW accuracy
\cite{Denner:2015yca,Denner:2016jyo,Denner:2016wet,Denner:2017kzu,%
Biedermann:2017bss,Biedermann:2016yds,Denner:2019tmn}.
For the purpose of this work, we have employed a private version of
\recola that enables the separation of weak-boson polarizations in
resonant SM amplitudes.

As a last comment of this section, we remark that the perturbative
order which we consider in this work is not state-of-the-art, as
we are neglecting both NNLO QCD and NLO EW corrections. However, the
leading radiative corrections are represented by the NLO QCD ones,
which are combined here with the results for the loop-induced $\Pg\Pg$
partonic process. In general, the impact of NLO EW corrections is
expected to be smaller than the one of QCD corrections. The results for $\PW^\pm
\PZ$ production in \rf{Baglio:2019nmc} show that angular distributions and
polarization observables are mildly modified by NLO EW corrections for what
concerns the $\PW$~boson in an inclusive setup.  We therefore expect
similar results in $\PW$-pair production, at least in the absence of
lepton cuts. However, we leave the treatment of NLO EW corrections to
future work.

\subsection{Input parameters and selection cuts}\label{sub:set}
We investigate the process  $\ppenmn$  for a proton--proton CM energy of
$13\TeV$ assuming SM dynamics.  At LO,
this process receives contributions from $\qqb$ initial
states only. The LO [$\mc O (\alpha^4)$] and NLO QCD [$\mc O (\as
\alpha^4)$] predictions are computed in the five-flavour scheme. We
choose the following on-shell EW vector-boson masses and widths
\cite{Tanabashi:2018oca},
\begin{alignat}{2}\label{smparqq}
 \Mwo &= 80.3790 \GeV,&\qquad \Gwo &= 2.0850\GeV, \notag\\
 \Mzo &= 91.1876 \GeV,&\qquad \Gzo &= 2.4952\GeV, 
\end{alignat}
which are converted into the corresponding pole values by means of \cite{Bardin:1988xt},
\beq
        M_V = \frac{\MVOS}{\sqrt{1+(\GVOS/\MVOS)^2}}\,,\qquad  
\Gamma_V = \frac{\GVOS}{\sqrt{1+(\GVOS/\MVOS)^2}}.
\eeq

Further SM parameters are chosen as 
\begin{alignat}{2}
\label{smpar2}
 \MH &= 125\GeV,&\qquad \GH &= 0.00407\GeV, \notag\\
 \Mt &= 173 \GeV,&\qquad \Gt &= 0\GeV, \notag\\
 \GF &= 1.16638\cdot10^{-5} \GeV^{-2}.
\end{alignat}
While the Higgs and top parameters are irrelevant for the
quark-induced processes, they enter the calculation of the
gluon-induced channel. We consider massless bottom quarks. Even if we
work in the five-flavour scheme, the ${\Pb\bar{\Pb}}$-initiated
partonic channel is neglected, since it is strongly PDF suppressed
with respect to other quark flavours.  The parton distribution
functions (PDF) are passed to \mocanlo via the \texttt{LHAPDF6}
interface \cite{Buckley:2014ana}. We use \texttt{NNPDF3.1} PDFs
\cite{Ball:2017nwa} computed with $\as(\Mz)=0.118$
[\texttt{NNPDF31\_(n)lo\_as\_0118} for (N)LO].  The $\GF$~scheme is
employed for fixing the EW coupling, and the weak vector bosons
are treated in the complex-mass scheme
\cite{Denner:2000bj,Denner:2005fg,Denner:2006ic}.


We also present results for the loop-induced gluon-initiated partonic
processes $\ggenmn$.  For
this, we employ the same parameters described above, apart from the
b-quark mass which is now set to $M_\Pb = 4.7\GeV$. However, we keep
working in the five-flavour scheme. We use the same PDF choice as
for NLO QCD corrections to the quark-induced process.

In all computations the factorization and renormalization scales are
set to the $\PW$~pole mass, $\mu_{\rF}=\mu_{\rR}=\Mw$.

We consider two different sets of selection cuts. To validate the
polarized distributions we use a first setup (labelled
\emph{inclusive}) that only involves a technical cut on the 
charged-lepton transverse momentum, $\pt^\ell> 0.01 \GeV$, whose effects on
the results are completely negligible, and a jet veto on additional
jets with $\ptj>35\GeV,|\eta_{\Pj}|<4.5$. We then consider a second
setup (labelled \emph{fiducial}) that mimics the fiducial signal
region defined in a recent ATLAS measurement \cite{Aaboud:2019nkz}:
\begin{itemize}
\item minimum transverse momentum of the charged leptons, $\ptl>27\GeV$;
\item maximum rapidity of the charged leptons, $|\eta_\ell|<2.5$;
\item minimum missing transverse momentum, $\ptmiss> 20\GeV$;
\item the same jet veto (no jets with $\ptj>35\GeV,|\eta_{\Pj}|<4.5$) as in the inclusive setup;
\item minimum invariant mass of the charged lepton-pair system,
  $M_{\Pe^+\mu^-}>55\GeV$.
\end{itemize}
The last invariant-mass cut is applied to reduce the Higgs background
down to approximately $1\%$ of the total $\PW\PW$-production
cross-section. This is important mostly in the study of the
gluon-induced partonic process, for which in any case we exclude the
Higgs peak region by imposing $M_{2\ell 2\nu}>130\GeV$. Note that in
the ATLAS paper \cite{Aaboud:2019nkz} a transverse momentum cut
($p_{\rT,\Pe^+\mu^-}>30\GeV$) is also applied to the
charged lepton pair to suppress the Drell--Yan background. However, we
do not use this cut in the following, since it is motivated by
experimental mis-reconstruction of the final state.  The
top-production background is suppressed in the ATLAS analysis by means
of a $\Pb$-jet veto (no ${\Pb}$ jets with $\ptb>20\GeV$ and
$|\eta_\Pb|<2.5$). In our discussion we assume a
perfect $\Pb$-jet veto for simplicity.

\section{Results}
In this section, we present phenomenological results for polarized
signals in $\PW$-pair production.
We have investigated both singly-polarized and doubly-polarized
configurations, since the experimental interest lies both in the
single-boson polarization fractions and in the extraction of the
doubly-longitudinal cross-section.  Na\"ively one could expect
that the doubly-polarized cross-sections are directly related to the
singly-polarized ones.  This statement is wrong, as the two
$\PW$-boson spin states are strongly correlated in di-boson
kinematics. This is due to the absence of additional jets in the final
state and to the constrained angular momentum balance of the initial
state that features two spin-1/2 particles in the leading partonic
channel.  Such a correlation is weaker if the weak-boson pair is
produced with additional jets \cite{Ballestrero:2017bxn}.  However,
even in that case the zero-correlation hypothesis is definitely not
realistic.

In the following, we also evaluate the contributions of the
non-resonant irreducible background and effects of interferences among
polarizations at the level of total and differential cross-sections.
The non-resonant background is defined as the difference between
results based on full matrix elements (full, for simplicity) and the
unpolarized results computed with the DPA as described in
\sect{sub:def}. Their size is \emph{a priori} expected to be of the
order of the intrinsic error of the DPA [$\mc O(\alpha)$].

Interferences among different polarization states are estimated as the
difference between the unpolarized DPA results and the sum of
polarized DPA ones [see \eqn{eq:interfdef}].
Such a sum runs over two singly-polarized contributions, \ie T and L,
or four doubly-polarized ones, \ie TT, TL, LT, LL states:
in the
former case, interferences are expected to be slightly smaller than in
the latter one, as the interferences corresponding to the unpolarized
bosons are implicitly included in the calculation. Even if, in
general, interferences are expected to vanish in the absence of lepton
cuts, there can be non-trivial effects due to the correlations of the
polarizations of the two produced bosons in the doubly-polarized case
and in distributions in the singly-polarized case.

We are going to present results at LO and NLO QCD for the leading
$\qqb$ channel and combine them with LO results for the $\Pg\Pg$
channel.

The loop-induced production of $\PW^+\PW^-$ pairs with two
initial-state gluons gives a contribution at $\mc O(\as^2\alpha^4)$.
Although it is formally part of the NNLO QCD corrections to $\PW$-pair
production, it is enhanced by the gluon luminosity in the proton. It
can be computed independently as it only involves ultraviolet- and
infrared-finite amplitudes.  We note that for massless quarks and
leptons only doubly-$\PW$-resonant diagrams contribute to the
gluon-induced process \cite{Binoth:2006mf}.  Among them, only box
diagrams and one triangle Higgs-exchange diagram are non-vanishing.
The triangle diagram, proportional to the top-quark Yukawa coupling,
is negligible if the Higgs-mass region is cut away. As a consequence,
the DPA reproduces the full result much better than in the
quark-induced channels.

For the \Pg\Pg~channel, we exclude the Higgs-resonance peak by an
invariant-mass cut $M_{2\ell2\nu}>130\GeV$. This reduces the
cross-section for $\ggenmn$ by $5\%$ relative to the one computed from
the complete invariant-mass spectrum. Given that the gluon-induced
contribution is roughly $7\%$ of the NLO QCD total cross-section for
$\qqb$, the Higgs signal accounts for a few permille in the combined
cross-section.

Before presenting results, we remind the reader that we consider a
jet veto (no jets in the region $\ptj>35\GeV,|\eta_{\Pj}|<4.5$).
This selection is applied in $\PW^+\PW^-$ production to suppress large
contributions from additional QCD radiation.  If such a veto is not
required, the results (both polarized and unpolarized) change
noticeably, as the additional jet recoils against the $\PW^+\PW^-$
system. In particular, avoiding the jet veto leads to a substantial improvement of the
quality of the DPA at NLO QCD, as doubly-resonant diagrams contribute
also in regions where they would be excluded (\eg large missing
transverse momentum) if the jet~veto was applied.
Nonetheless, dropping the jet veto would mean that
$\PW^+\PW^-{\Pj}$ production (LO accurate) is not suppressed anymore with
respect to $\PW^+\PW^-$ production. Therefore the polarized
predictions would not pertain exactly to di-boson production.
In all the following we understand the jet veto described in \sect{sub:set}
to be applied.

  A further comment should be made on the jet veto.
  The results we are going to show concern fixed-order predictions
  in perturbative QCD. It is well known
  \cite{Monni:2014zra,Becher:2014aya} that applying a jet-veto in
  di-boson production leads to an enhancement of higher-order
  corrections due to large logarithms stemming from the ratio between
  the di-boson invariant mass ($\gtrsim 2\Mw$) and the jet-veto scale.
  While this affects total cross-sections and shapes of distributions,
  we do not expect that the resummation of jet-veto logarithms would
  sizeably affect the polarization fractions. This is supported by the
  fact that polarization fractions exhibit a very mild dependence on
  the QCD scale, as shown below. Moreover, for di-boson production with
  leptonic decays, the QCD corrections are related only to the initial
  state, while the polarization dependence is tied to the final
  state.

\subsection{Inclusive phase-space region}\label{sub:inc}
In this section we present results obtained in the inclusive setup.
For \PW-pair production, the final state only involves the decay
products of the two $\PW$ bosons, which are produced almost
back-to-back, up to an additional jet with small $\pt$ at NLO.
Therefore, the kinematic variables of decay products are more strongly
correlated than in other multi-boson signatures, such as in vector-boson
scattering. This affects the polarized signals leading to interesting
results that could be na\"ively considered as unexpected.

In \tab{table:sigmainclNLOvetoed} we show the total cross-sections (in
fb) for all relevant polarizations, including both singly- and
doubly-polarized results. The numerical errors on the central values
are indicated in parentheses. The percentage scale uncertainties,
extracted with 7-point variations around the central scale, are
provided in superscripts and subscripts.
\begin{table}
\begin{center}
\renewcommand{\arraystretch}{1.3}
\begin{tabular}{C{3.8cm}C{2.3cm}C{2.3cm}C{1.7cm}C{1.7cm}}%
\hline %
\cellcolor{blue!9}   & \cellcolor{blue!9}  LO  & \cellcolor{blue!9} {NLO QCD}       & \cellcolor{blue!9} {$K$-factor}&   \cellcolor{blue!9} {$\Delta_{\Pg\Pg}$}  \\
\hline
 full   & $ 871.4(4)^{+4.2\%}_{-5.1\%}$   &  $932.0(9)^{+1.8\%}_{-2.3\%}$                         & 1.07 &  1.05   \\
unpolarized (DPA)   &   $859.1(2)^{+4.2\%}_{-5.1\%}$   &  $920.7(5)^{+1.9\%}_{-2.4\%}$            & 1.07 &  1.05  \\
\hline
 $\PW^+_{\rL}\PW^{-}_{\rm unpol}$ (DPA)   &  $224.0(1)^{+5.0\%}_{-6.3\%}$    &  $249.2(2)^{+2.0\%}_{-2.5\%}$ & 1.11 &  1.03  \\
 $\PW^+_\rT\PW^{-}_{\rm unpol}$ (DPA)    &  $635.0(1)^{+4.0\%}_{-4.8\%}$   &  $671.4(4)^{+1.8\%}_{-2.2\%}$ & 1.06 &  1.06  \\
\hline
 $\PW^+_{\rL}\PW^-_{\rL}$ (DPA)   &   $16.22(1)^{+5.2\%}_{-6.0\%} $   &  $25.19(2)^{+2.4\%}_{-3.3\%}$         & 1.55 &  1.08  \\
 $\PW^+_{\rL}\PW^-_\rT$ (DPA)    &  $207.8(1)^{+5.0\%}_{-6.0\%}$    &  $224.0(1)^{+2.0\%}_{-2.6\%}$         & 1.08 &  1.03  \\
 $\PW^+_\rT\PW^-_{\rL}$ (DPA)    &  $253.9(1)^{+4.8\%}_{-5.8\%}$  &  $266.3(2)^{+2.0\%}_{-2.5\%}$           & 1.05 &  1.02  \\
 $\PW^+_\rT\PW^-_\rT$ (DPA)    &  $381.1(1)^{+3.3\%}_{-4.1\%}$    &  $404.9(2)^{+1.6\%}_{-2.0\%}$         & 1.06 &  1.08  \\
\hline
\end{tabular}
\end{center}
\caption{Total cross-sections (in fb) in the inclusive setup for the
  unpolarized, singly-polarized and doubly-polarized $\PW^+\PW^-$
  production at the LHC. 
Uncertainties are computed with 7-point scale variations. $K$-factors
are computed as ratios of NLO QCD over LO integrated
cross-sections. The contribution of the gluon-induced channel is shown
relative to the NLO QCD results for $\qqb$.} 
\label{table:sigmainclNLOvetoed}
\end{table}
Note that the DPA cross-sections are identically zero for
$M_{2\ell2\nu}<2\Mw$ by definition.  The contribution to the full
cross-section of the region below $2\Mw$ is merely 1.3\%.  As a general
comment, the transverse polarization strongly dominates over the
longitudinal one, as in most multi-boson production processes
\cite{Bern:2011ie,Stirling:2012zt}. This implies that $K$-factor,
scale variations, and enhancement due to the \Pg\Pg~channel for
transverse $\PW$~bosons are very similar to the unpolarized case.
Results with at least one longitudinal boson show slightly larger
$K$-factors compared to the transverse ones: 1.11 for the
singly-longitudinal (dominated by the transverse component of the
unpolarized boson), 1.55 for the LL, despite the
application of the jet veto.  After combining with the gluon-induced
channel, the unpolarized cross-section is enhanced by 5\%, the
LL and TT cross-sections
by $8\%$, and the mixed ones by just
$2$--$3\%$. The difference between the two mixed doubly-polarized
cross-sections results from the different angular momentum balance
in the u-type and d-type quark-initiated partonic channels.  Since no
cuts are imposed on the leptons, the sum of singly- or
doubly-polarized cross-sections is identical to the unpolarized DPA
cross-section within integration errors. This is further confirmed by the analysis
of differential distributions.

As a validation of our definition of polarized vector bosons, we
consider the distributions in the decay angles of leptons
$\theta^*_{\ell}$ and $\phi^*_{\ell}$ computed in the corresponding
$\PW$-boson rest frame.  The differential distributions in these
variables directly reflect the polarization modes of the decayed
vector boson. An ambiguity is related to the reference axis with
respect to which the leptonic angular variables $\theta^*_\ell$,
and $\phi^*_\ell$ in the weak-boson rest frame are defined.  

Using the
so-called helicity coordinate system \cite{Bern:2011ie}, we choose as
reference axis the direction of the $\PW$ boson in the laboratory
frame. With this choice, the polarization vectors in the laboratory
frame and the helicity frame are directly related by a boost along the
reference axis, and the dependence on 
$\theta^*_\ell$ and $\phi^*_\ell$ directly reflects the polarizations
in the laboratory frame. This would not be the case if the reference
axis was chosen as the direction of the $\PW$ boson in the CMS frame
of the W-boson pair.

\begin{figure}
  \centering
 \subfigure[Cosine of theta angle of $\Pe^+$ in the $\PW^+$ CM frame.\label{costhetastarINC}]{\includegraphics[scale=0.36]{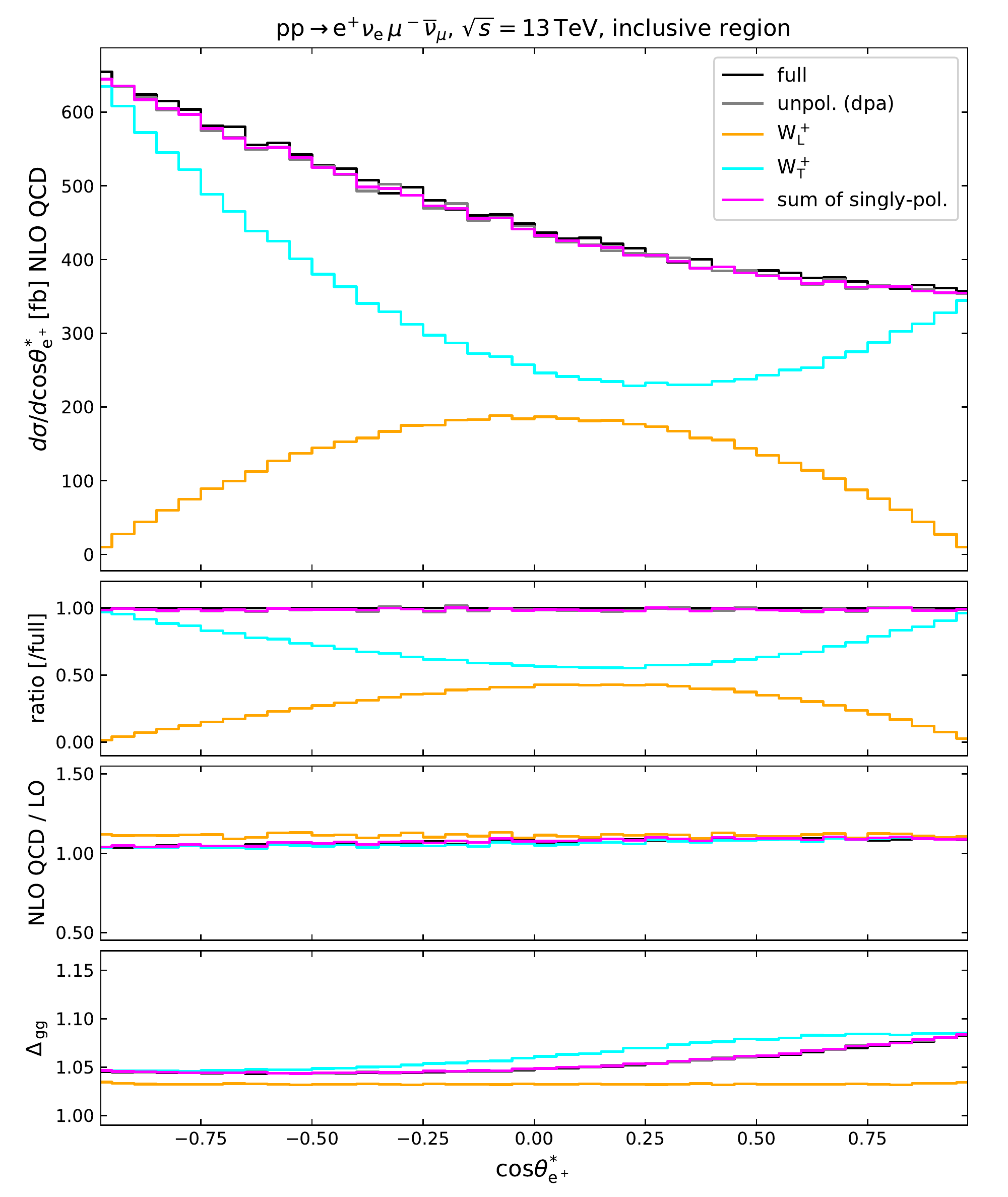}}
 \subfigure[Azimuthal angle of $\Pe^+$ in the $\PW^+$ CM frame.\label{phistarINC}]{\includegraphics[scale=0.36]{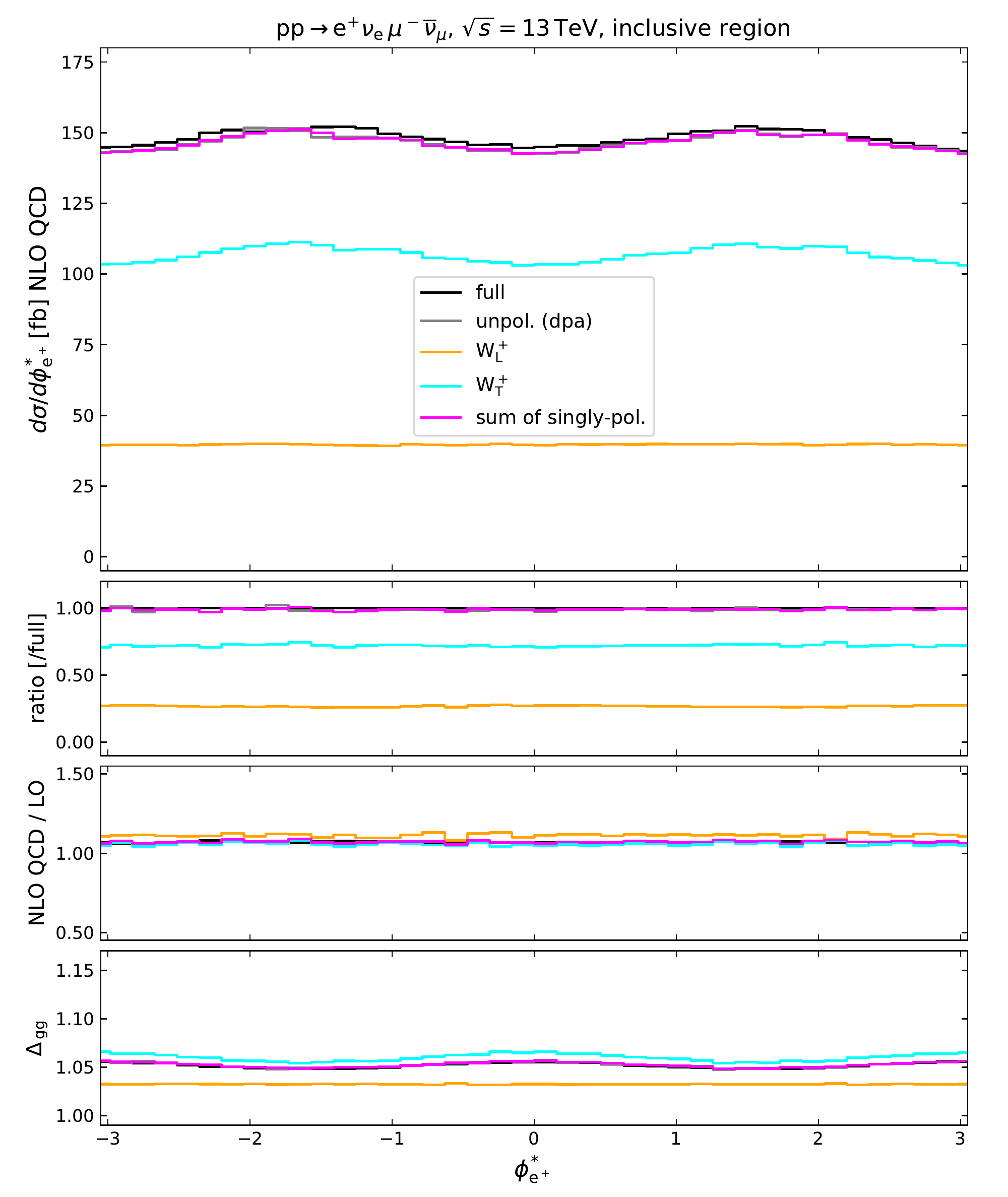}}  
 \caption{Distributions in the positron angular variables
   $\cos\theta^*_\Pe$ and $\phi^*_\Pe$ defined in the $\PW^+$ rest frame
   in the inclusive region. Singly-polarized and unpolarized results
   are shown. From top down: NLO QCD differential cross-sections for
   the $\qqb$ channel, ratios over the full (unpolarized), $K$-factors
   (NLO QCD/ LO), and enhancement of the NLO QCD cross-sections due to
   the $\Pg\Pg$ channel. 
 }
 \label{stardistribINC1}
\end{figure}

We consider singly-polarized distributions for the $\PW^+$ boson
decaying to $\Pe^+\nu_\Pe$, but analogous results can be obtained for
the $\PW^-$ boson.  We stress that, while being sensitive to the
polarization state of the weak bosons, these angular variables require
the full reconstruction of both $\PW$~bosons separately, which is
impossible with two neutrinos in the final state.
However, they are relevant to validate the
definition of polarized cross-sections in Monte Carlo simulations.
In the corresponding differential distributions, shown in
\fig{stardistribINC1}, we observe that the interferences are
compatible with zero in the inclusive setup, as expected.  The
$K$-factor is practically 
constant for both the polarized and the
unpolarized configurations.

The most important reason to study distributions in $\theta^*_\Pe$ and
$\phi^*_\Pe$ lies in the possibility of extracting analytically the
polarization fractions from the unpolarized distribution.  At LO in
the EW coupling, the differential cross-section for the production of
an unpolarized $\PW^+$~boson reads
\begin{align}
  \frac{\rd\sigma}{\rd\cos\theta^*_\Pe \,\rd\phi^*_\Pe \,\rd X}
  ={}&\frac{\rd\sigma}{\rd X}\frac{3}{16\pi} 
\biggl[(1+\cos^2{\theta^*_\Pe})+({A_0}/{2})(1-3\cos^2{\theta^*_\Pe})+A_1 \sin2\theta^*_\Pe\,\cos\phi^*_\Pe \notag\\
    & \hspace{1.2cm}+(A_2/2)\sin^2\theta^*_\Pe\,\cos2\phi^*_\Pe  
    +A_3\sin\theta^*_\Pe\,\cos\phi^*_\Pe+ A_4\cos\theta^*_\Pe \notag\\
    &\hspace{1.2cm}+ A_5\sin^2\theta^*_\Pe\,\sin2\phi^*_\Pe + A_6\sin2\theta^*_\Pe\,\sin\phi^*_\Pe + A_7\sin\theta^*_\Pe\,\sin\phi^*_\Pe\biggr]\,,
\label{eq:mastereq}
\end{align}
where $X$ is a generic (set of) kinematic variable(s), independent of
$\theta^*_\Pe$ and $\phi^*_\Pe$. The  coefficients $A_i$ represent
scalar quantities that are related to the polarization of the produced
$\PW$~boson, and depend on the variable $X$.
If the full $\phi^*_\Pe$ range is accessible, which is the case if no
cuts are imposed on the positron and the related neutrino, the
interferences vanish upon integrating out the azimuthal angle 
leaving a simple functional dependence on $\theta^*_\Pe$:
\begin{align}
  \frac{\rd\sigma}{\rd\cos\theta^*_\Pe \,\rd X}
  ={}&\frac{\rd\sigma}{\rd X}\frac{3}{8} 
\biggl[(1+\cos^2{\theta^*_\Pe})+({A_0}/{2})(1-3\cos^2{\theta^*_\Pe})
    + A_4\cos\theta^*_\Pe\biggr]\,\notag\\
  ={}& \frac{\rd\sigma}{\rd X}\frac{3}{8} \biggl[2f_{\rL} \sin^2{\theta^*_\Pe}+ f_{-}(1-\cos{\theta^*_\Pe})^2+f_{+}(1+\cos{\theta^*_\Pe})^2\biggr]\,,
\label{eq:mastereq_2}
\end{align}
where the polarization fractions $f_i$ are related to $A_i$ by simple
linear combinations, and are such that $f_{\rL}+f_{+}+f_{-}=1$.  This
expression can be used to extract polarization fractions for
processes which are dominated by vector-boson resonances, like
$\PW$-pair production, provided that the decay products 
can be uniquely identified.

We have extracted the polarization fractions from the unpolarized DPA
distribution in $\cos\theta^*_\Pe$ by means of suitable projections on
\refeq{eq:mastereq_2} in the same fashion as in
\rf{Ballestrero:2017bxn} and combined them into $f_{\rL}$ and $f_{\rT}
= f_{+}+f_{-}$.
These are compared with polarization fractions that are obtained as
ratios of the polarized cross-sections (computed with the Monte Carlo)
over the unpolarized DPA one. The agreement is almost perfect, as can
be seen in \tab{table:sigma4V_fractions}.
\begin{table}
\begin{center}
\renewcommand{\arraystretch}{1.3}
\begin{tabular}{C{3.8cm}|C{2.3cm}C{2.3cm}|C{2.3cm}C{2.3cm}}
\hline
\cellcolor{blue!9}  & \multicolumn{2}{c|}{\cellcolor{blue!9} \bf LO } & \multicolumn{2}{c}{\cellcolor{blue!9} \bf NLO QCD} \\
\hline
non-resonant backgr.  & \multicolumn{2}{c|}{$0.0142(5)^{+0.0002}_{-0.0003}$ } & \multicolumn{2}{c}{$0.0111(9)^{+0.0003}_{-0.0003}$ }\\
interferences  & \multicolumn{2}{c|}{$< 10^{-4}$ } & \multicolumn{2}{c}{$< 10^{-4}$}\\
\hline
& \bf MC  & \bf analytic & \bf MC & {\bf analytic} \\
 $\PW^+_{\rL}\PW^{-}_{\rm unpol}$ (DPA)   &  $0.261(1)^{+0.002}_{-0.002}$   &  $0.260(3)^{+0.002}_{-0.003}$  & $0.271(1) ^{+0.001}_{-0.001}$  & $0.272(3)^{+0.001}_{-0.001}$ \\
 $\PW^+_{\rT}\PW^{-}_{\rm unpol}$ (DPA)    &  $0.739(3)^{+0.003}_{-0.002}$  &  $0.740(5)^{+0.002}_{-0.002}$ & $0.729(3) ^{+0.002}_{-0.001}$  &  $0.728(6)^{+0.001}_{-0.001}$\\
\hline
\end{tabular}
\end{center}
\caption{Fractions for a polarized $\PW^+$ boson produced in association
  with an unpolarized $\PW^-$~boson in the inclusive setup.
  Uncertainties are computed with 7-point scale variations. The
  polarization fractions from Monte Carlo ({\bf MC}) are obtained by
  taking the ratio of polarized total cross-sections over the
  unpolarized one in the DPA. The same holds for contributions of the  
  non-resonant irreducible background and interferences. The {\bf
    analytic} polarization
  fractions are obtained by suitable projections of the unpolarized DPA
  cross-section.} 
\label{table:sigma4V_fractions}
\end{table}
Note that the results extracted from unpolarized angular distributions
(analytic) agree perfectly with Monte Carlo predictions not only at
the normalization level (total cross-sections) but also for the shapes of
singly-polarized distributions in $\cos\theta^*_\Pe$.  The LO
longitudinal polarization fraction experiences a $1\%$ enhancement due to NLO
corrections, which is balanced by a corresponding decrease in
the transverse fraction. We also observe that the obtained
polarization fractions are very stable against scale variations both
at LO and NLO QCD.
The polarization fractions presented in \tab{table:sigma4V_fractions} concern the
complete inclusive phase-space region, but a very good agreement
($<1\%$) is found even in specific ranges of the $\PW^+$
transverse momentum and rapidity.

As can be deduced from the asymmetry of the distribution for
transverse $\PW^+$~boson in \fig{costhetastarINC} combined with
\eqn{eq:mastereq_2}, the left-handed polarization is almost twice the
right-handed one in the $\qqb$ channel ($f_{+}/f_{-}\approx
0.52$), in good agreement with the corresponding result of Table~2 in
\rf{Stirling:2012zt}. This asymmetry is due to the fact that the
process is $\qqb$ initiated, so that the $\PW$~bosons are preferably
generated with left-handed helicity. As the produced bosons have
preferably small $\pt$, angular-momentum arguments \cite{Bern:2011ie}
imply that the $\PW^+$~boson originating from
${\Pu\bar{\Pu}}$/${\Pc\bar{\Pc}}$ annihilation is mostly left handed
($f_{-} \approx 0.67$, $f_{+} \approx 0.11$), while in
${\Pd\bar{\Pd}}$/${\Ps\bar{\Ps}}$ it is mostly right handed ($f_{-}
\approx 0.21$, $f_{+} \approx 0.47$). Weighting the polarization fractions with
the relative PDF factors between the two quark--antiquark channels,
one recovers $f_{+}/f_{-}\approx 0.52$.  On the contrary, the
loop-induced channel gives perfect left--right symmetric distributions
for transverse $\PW^+$~bosons, as expected for zero-charge, spin-1
massless particles in the initial state.

As seen in \fig{phistarINC}, the $\phi^*_\Pe$ distribution for the
longitudinal polarization is flat, as expected from the decay amplitudes,
 while the transverse one receives a $\cos 2\phi^*_\Pe$
modulation, as the interference between the left- and right-handed
modes gives non-vanishing $\phi^*_\Pe$-dependent terms. This is exactly
the origin of the $\phi^{*}_\Pe$ dependence in \eqn{eq:mastereq}. We
have checked that simulating the production of left-handed or
right-handed \PW~bosons separately gives flat distributions, as
the $\phi^*_\Pe$ dependence disappears in the squared amplitudes
($\mc A_{\pm} \propto \re^{\pm \ri{\phi^*_\Pe}}$). Note that this
argument is no longer true in the presence of lepton cuts.  An
interesting aspect is that in the gluon-induced process the spin-1
nature of the incoming partons also gives a $\cos 2\phi^{*}_\Pe$
modulation but with opposite sign with respect to the quark-induced
channel.

So far, we have provided a number of arguments that prove the quality
of polarization separation at the amplitude level and in the Monte
Carlo simulation.  This has been validated not only for singly-polarized
signals but also for the doubly-polarized ones. Considering the
$\cos\theta^*_\mu$ distribution for a longitudinal $\PW^+$ and an
unpolarized $\PW^-$~boson, we have extracted via suitable projections
\cite{Ballestrero:2017bxn} the doubly-polarized distributions (LL, LT)
and found them to agree perfectly with the results directly simulated
with the Monte Carlo.  Analogously, we have extracted the TL and TT
components from the transverse--unpolarized $\cos\theta^*_\mu$
distributions.  These checks further confirm that the proposed
definition of polarized signals is very well behaved.

Furthermore, the integrated cross-sections and the distributions in
$\cos\theta_\Pe^*$ presented so far confirm that the interferences
among polarization modes vanish upon integration over the azimuthal
decay angles in the absence of lepton cuts. This holds for most
of the kinematic observables that can be reconstructed at the LHC.

\begin{figure}
  \centering
 \subfigure[Missing transverse momentum.\label{ptmiss}]{\includegraphics[scale=0.36]{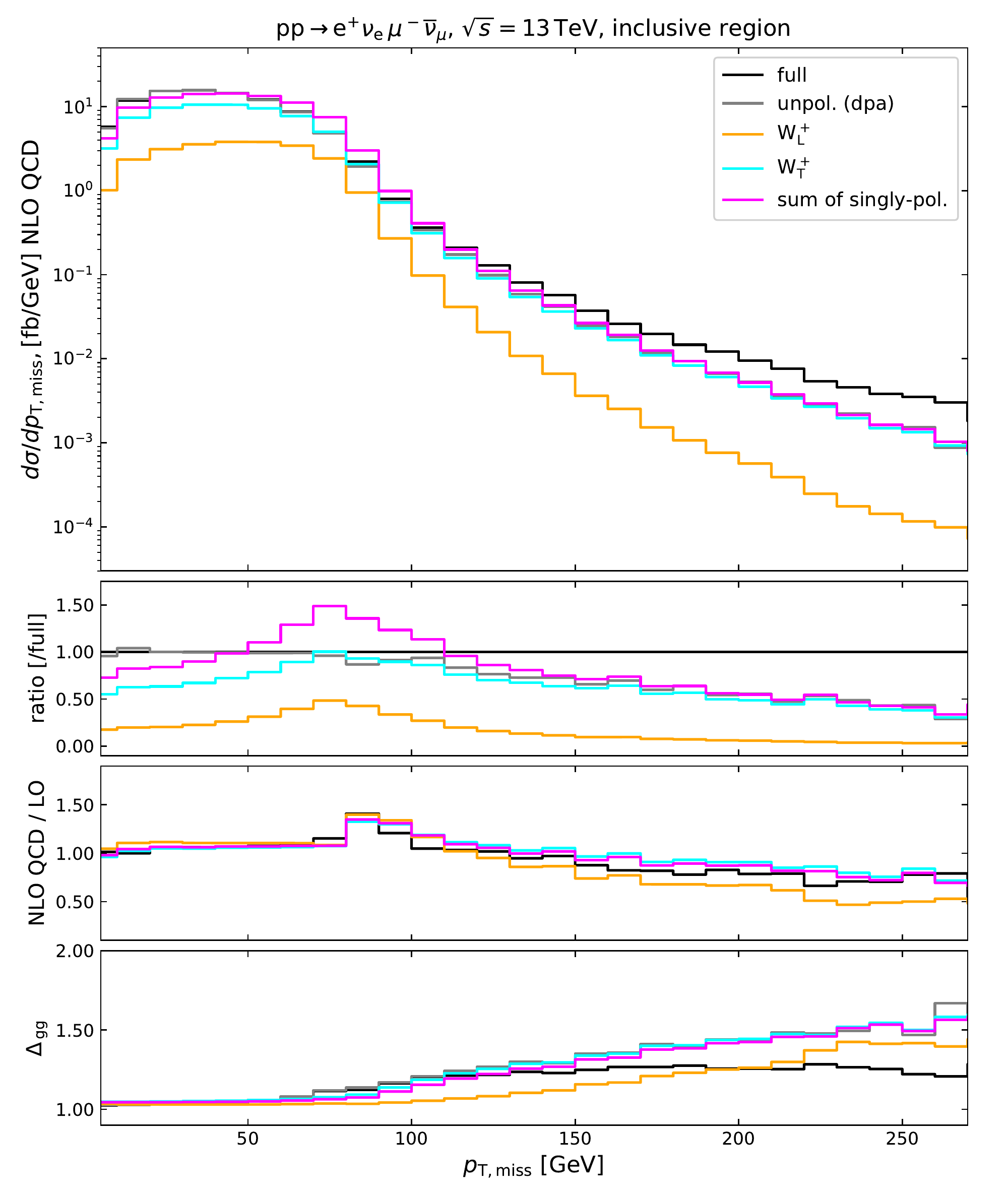}}
 \subfigure[Azimuthal separation between $\Pe^+$ and $\mu^-$.\label{azimuthll}]{\includegraphics[scale=0.36]{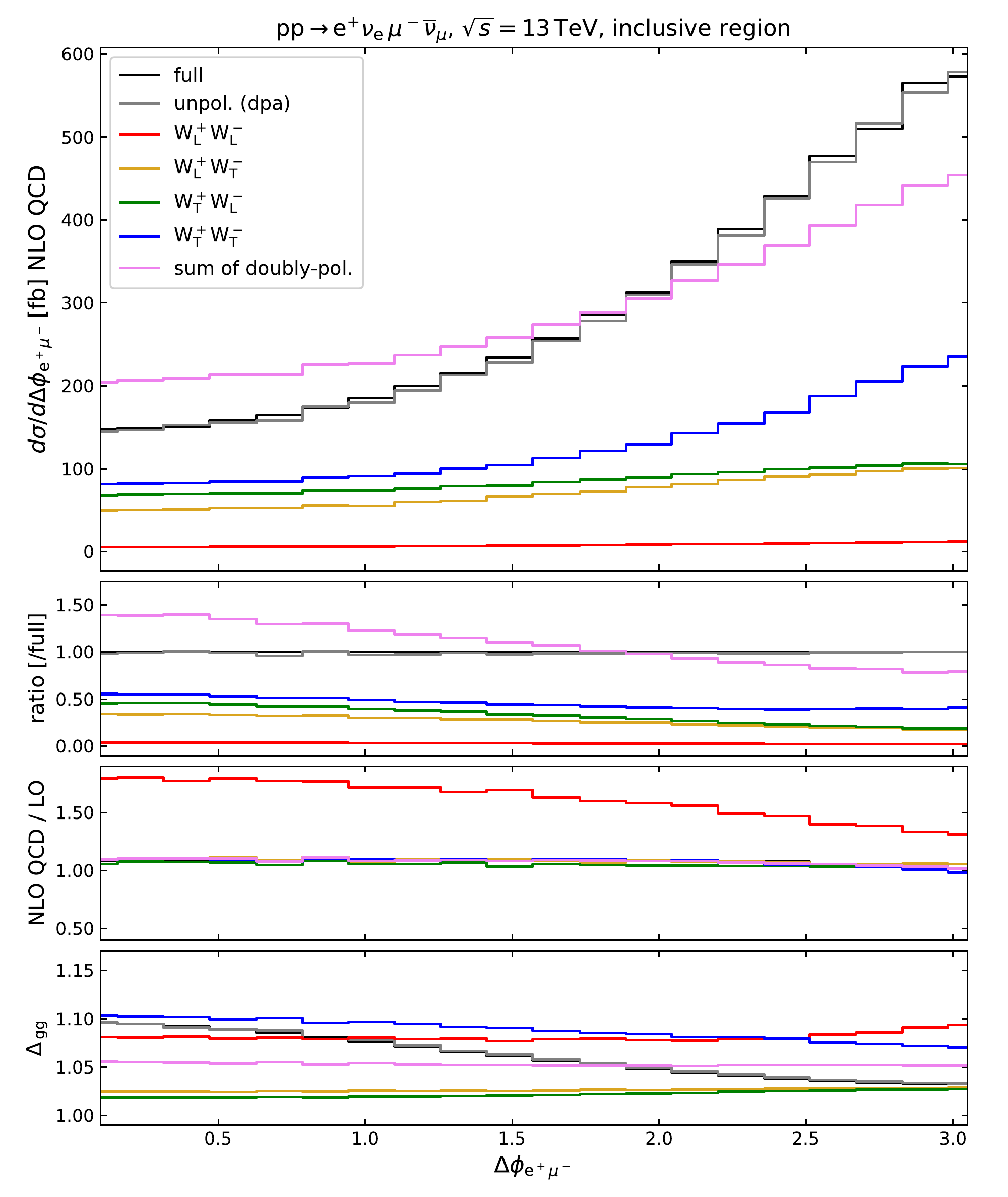}}  
 \caption{Distributions in the missing transverse momentum and in the
   azimuthal separation between the two charged leptons in the
   inclusive region. Singly-polarized results are shown in
   \fig{ptmiss}, doubly-polarized ones in \fig{azimuthll}.  Same
   subplot structure as in
   \fig{stardistribINC1}.}\label{stardistribINC2}.
\end{figure}

However, for some observables, even in the inclusive setup, the
polarization separation does not provide predictions that can easily
be interpreted as polarized predictions.  A first example is given by
the distribution in the missing transverse momentum, shown for the
polarized and unpolarized cases in \fig{ptmiss}.
It is well known that in kinematic regions that are not dominated by
doubly-resonant diagrams the DPA describes the full
computation badly. This is the case for the region of large missing
transverse momentum, where the doubly-resonant contributions are
suppressed and the singly-resonant diagrams (dropped in the DPA) give
a relevant contribution to the full calculation already at Born level
\cite{Biedermann:2016guo}. This translates into a suppression of
polarized distributions, as can be seen for all DPA curves in
\fig{ptmiss} for $\ptmiss > 100\GeV$. The discrepancy between the
unpolarized DPA and full results reaches $-50\%$ already at moderate
values $\ptmiss\approx 200\GeV$. 

Furthermore, even in the region where the DPA behaves well ($\ptmiss
\lesssim 100\GeV$), the sum of singly-polarized distributions does not
reproduce the unpolarized DPA one, pointing out large interferences
between the transverse and longitudinal modes.  In fact, selecting the
missing transverse momentum in a certain range imposes restrictions on
the lepton kinematics, specifically on a variable that is not directly
related to a single $\PW$~boson, and impedes the cancellation of
interferences. A similar situation is observed also for the
distributions in the transverse momentum of the positron--muon pairs.
This effect has been found also in $\PW^+\PW^-$ scattering
\cite{Ballestrero:2017bxn}.

The above comments apply to both LO and NLO QCD results.  The
$K$-factor is similar for the various polarized and unpolarized
distributions: only the longitudinal polarization features a somewhat
smaller $K$-factor in the large $\ptmiss$ regime.  In the
gluon-induced channel, the doubly-resonant diagrams are dominant even
at large $\ptmiss$, leading to a better description of the full matrix
elements and a better modelling of the polarized signals.

Another interesting variable that discloses some surprises in the
inclusive setup is the azimuthal separation between the two charged
leptons shown in \fig{azimuthll}.  In this case the unpolarized DPA
distribution describes the full result (with at most 2\% deviation)
very well.  However, the sum of doubly-polarized signals is far from
the unpolarized one, \ie large interferences characterize this
observable. They amount to $+40\%$ for $\Delta\phi_{\Pe^+\mu^-}\approx
0$ and $-30\%$ for $\Delta\phi_{\Pe^+\mu^-}\approx \pi$. The same
happens for singly-polarized distributions.  Note that such an effect
can only result from interferences between the longitudinal and the
transverse mode, as the left--right interference is already accounted
for in the definition of the cross-sections for transverse
polarization.  A similar situation has been found in the $\PW^+\PW^-$
decay of a Higgs boson produced in gluon~fusion \cite{MainaHiggs},
while it is absent in vector-boson scattering
\cite{Ballestrero:2017bxn}.  Whereas in vector-boson scattering the
two $\PW$~bosons are produced mostly in the central region, in
di-boson production they feature back-to-back
kinematics inducing correlations between the decay angles of the two
charged leptons. This is supported by the fact that this large effect
is partially reduced upon omitting the jet veto in NLO QCD
corrections. 
Since such interferences are not found for
$\PZ\PZ$~production, as we checked numerically, they are apparently
enhanced in the back-to-back kinematics by the purely left-handed
nature of the $\PW$-boson coupling to leptons.

In the gluon-induced channel the interferences show an opposite
behaviour, being positive for $\Delta\phi_{\Pe^+\mu^-}\approx \pi$ and
negative close to 0. However, the combination of all
partonic processes features the same behaviour as the dominant
quark-induced process.

It is evident that the correlation between the polarizations of the
two $\PW$~bosons affects the polarized cross-sections in $\PW^+\PW^-$
production much more than in the presence of additional jets, such as in
$\PW^+\PW^-\Pj\Pj$ production. As a consequence, interferences can
appear even in the absence of cuts on single leptons, reducing the
number of variables that allow for an interpretation of unpolarized
distributions as a sum of the polarized ones.  Nevertheless, the
results of this section already show that defining accurately
polarized signals and accounting for interferences is definitely
needed to enable the correct extraction of polarized information from
LHC data.


\subsection{Fiducial phase-space region}\label{sub:fid}
Relying on the validation of our definition of polarized signals performed
in the inclusive setup, we are ready to present results in the
fiducial region targeting a realistic analysis of polarized
$\PW^+\PW^-$ production at the LHC.

The cuts on final-state leptons are expected to generate non-negligible
interferences both at the level of total cross-sections and in
differential distributions. This renders the analytic expressions in
\eqn{eq:mastereq_2} not valid anymore: their application in the
presence of realistic lepton cuts (as done in \rf{Baglio:2018rcu})
gives results that can be far from the actual
polarization structure of the process.

In analogy with \tab{table:sigmainclNLOvetoed}, we show in
\tab{table:sigmaleptNLOvetoed} singly- and doubly-polarized fiducial
cross-sections.
\begin{table}
\renewcommand{\arraystretch}{1.3}
\begin{center}
\begin{tabular}{C{3.8cm}|C{2.5cm}C{2.5cm}C{1.7cm}C{1.7cm}}
\hline
\cellcolor{blue!9}   & \cellcolor{blue!9}  LO  & \cellcolor{blue!9} {NLO QCD}       & \cellcolor{blue!9} { $K$-factor}&   \cellcolor{blue!9} { $\Delta_{\Pg\Pg}$}  \\
\hline
 full   & $ 202.02(3)^{+4.6\%}_{-5.5\%}$   &  $220.16(8)^{+1.8\%}_{-2.2\%}$                         & 1.09 &  1.06   \\
unpolarized (DPA)   &   $195.91(3)^{+4.7\%}_{-5.5\%}$   &  $214.48(9)^{+1.8\%}_{-2.2\%}$            & 1.09 &  1.06  \\
\hline
 $\PW^+_\rL\PW^-_{\rm unpol}$ (DPA)   &  $50.94(1)^{+5.5\%}_{-6.5\%}$    &  $57.42(4)^{+1.9\%}_{-2.6\%}$ & 1.13 &  1.04  \\
 $\PW^+_{\rT}\PW^-_{\rm unpol}$ (DPA)    &  $141.72(2)^{+4.3\%}_{-5.1\%}$   &  $152.84(9)^{+1.7\%}_{-2.1\%}$ & 1.08 &  1.07  \\
\hline
 $\PW^+_\rL\PW^-_\rL$ (DPA)   &   $6.653(1)^{+4.9\%}_{-5.8\%} $   &  $9.057(5)^{+2.9\%}_{-3.0\%}$         & 1.36 &  1.08  \\
 $\PW^+_\rL\PW^-_{\rT}$ (DPA)    &  $44.08(1)^{+5.6\%}_{-6.5\%}$    &  $48.24(4)^{+1.9\%}_{-2.5\%}$         & 1.09 &  1.04  \\
 $\PW^+_{\rT}\PW^-_\rL$ (DPA)    &  $50.19(1)^{+5.5\%}_{-6.4\%}$  &  $54.02(4)^{+1.9\%}_{-2.5\%}$           & 1.08 &  1.03  \\
 $\PW^+_{\rT}\PW^-_{\rT}$ (DPA)    &  $99.61(2)^{+3.7\%}_{-4.6\%}$    &  $106.20(7)^{+1.6\%}_{-1.9\%}$         & 1.07 &  1.09  \\
\hline
\end{tabular}
\end{center}
\caption{Fiducial cross-sections (in fb). Same observables as in \tab{table:sigmainclNLOvetoed}.}
\label{table:sigmaleptNLOvetoed}
\end{table}
At the integrated level, the results feature common aspects with those
obtained in the inclusive setup.  The contribution of NLO QCD
corrections is slightly enhanced by the lepton cuts both for the
unpolarized and for the singly-polarized case ($+2\%$). The
doubly-polarized cross-sections undergo a $+1\%$ increase of the
$K$-factors for those cross-sections involving at least one transverse
boson. On the contrary, the $K$-factor for the
LL cross-section is 12\%
smaller than in the inclusive setup but still much larger than for
other polarization combinations.  The combination of NLO QCD
results with those for the gluon-induced process leads to
an enhancement of $1\%$ relative to the inclusive setup for all
polarization combinations.
The transverse polarization is dominant both in the singly- and in the
doubly-polarized case.  However, even the LL
cross-section promises a reasonable number of events with the
luminosity accumulated during Run 2 of the LHC. This gives us confidence
that a detailed study of doubly-polarized signals will be performed in
this process.

Starting from the results of \tab{table:sigmaleptNLOvetoed}, we have
evaluated the polarization fractions in the singly-polarized case, as
well as the contributions of non-resonant irreducible background and
interferences for the fiducial cross-section (see
\tab{table:sigma4V_fractions_leptoncut}).
\begin{table}
\begin{center}
\renewcommand{\arraystretch}{1.3}
\begin{tabular}{C{4.4cm}|C{2.7cm}C{2.7cm}}
\hline
\cellcolor{blue!9}  &{\cellcolor{blue!9} \bf LO } & {\cellcolor{blue!9} \bf NLO QCD} \\
\hline
non-resonant background & {$0.0302(4)^{+0.0008}_{-0.0009}$ } & {$0.0258(7)^{+0.0003}_{-0.0002}$ }\\
interferences  & {$0.017(1)$ } & { $0.020(2)$ }\\
 $\PW^+_{\rL}\PW^{-}_{\rm unpol}$ (DPA)   &  $0.260(1)^{+0.002}_{-0.003}$  & $0.268(1)^{+0.001}_{-0.001}$ \\
 $\PW^+_{\rT}\PW^{-}_{\rm unpol}$ (DPA)   &  $0.723(2)^{+0.003}_{-0.003}$  &  $0.712(2)^{+0.001}_{-0.001}$ \\
\hline
\end{tabular}
\end{center}
\caption{Fractions for a polarized $\PW^+$~boson produced in
  association with an unpolarized $\PW^-$ boson in the fiducial
  region. Uncertainties are computed with 7-point scale variations. The
  polarization fractions are obtained by taking the ratio of polarized
  total cross-sections over the unpolarized one in the DPA. }
\label{table:sigma4V_fractions_leptoncut}
\end{table}
Similar to the inclusive setup, the polarization
fractions are very stable against QCD radiative corrections, featuring
again a $\pm 1\%$ modification between LO and NLO QCD and 
against scale variations. Furthermore, the polarization fractions in
the fiducial phase space are similar to those computed in the
inclusive setup.
 
In the fiducial region, the non-resonant background contributions are
slightly enhanced, as they account for $2.6\%$ of the full cross-section
at NLO QCD, which should be compared with the $1.1\%$ in the absence of
selection cuts.  The application of cuts on the lepton kinematics (in
particular on the transverse momentum of a single lepton) induces
non-vanishing interferences among polarization states, which amount to
$2\%$ of the full unpolarized result. This contribution is small but
non-negligible even at the integrated level. We show below that at the
differential level, the interference effects are strongly enhanced in
certain phase-space regions.

We now present polarized distributions for some selected kinematic
variables. We stress that only a limited number of them represents
measurable quantities at the LHC. However, studying non-measurable
quantities enables us to find possible similarities with other LHC
observables.

\begin{figure}
  \centering
  \subfigure[Cosine of polar angle of $\Pe^+$ in the $\PW^+$ CM frame.\label{costhstar_lep}]{\includegraphics[scale=0.36]{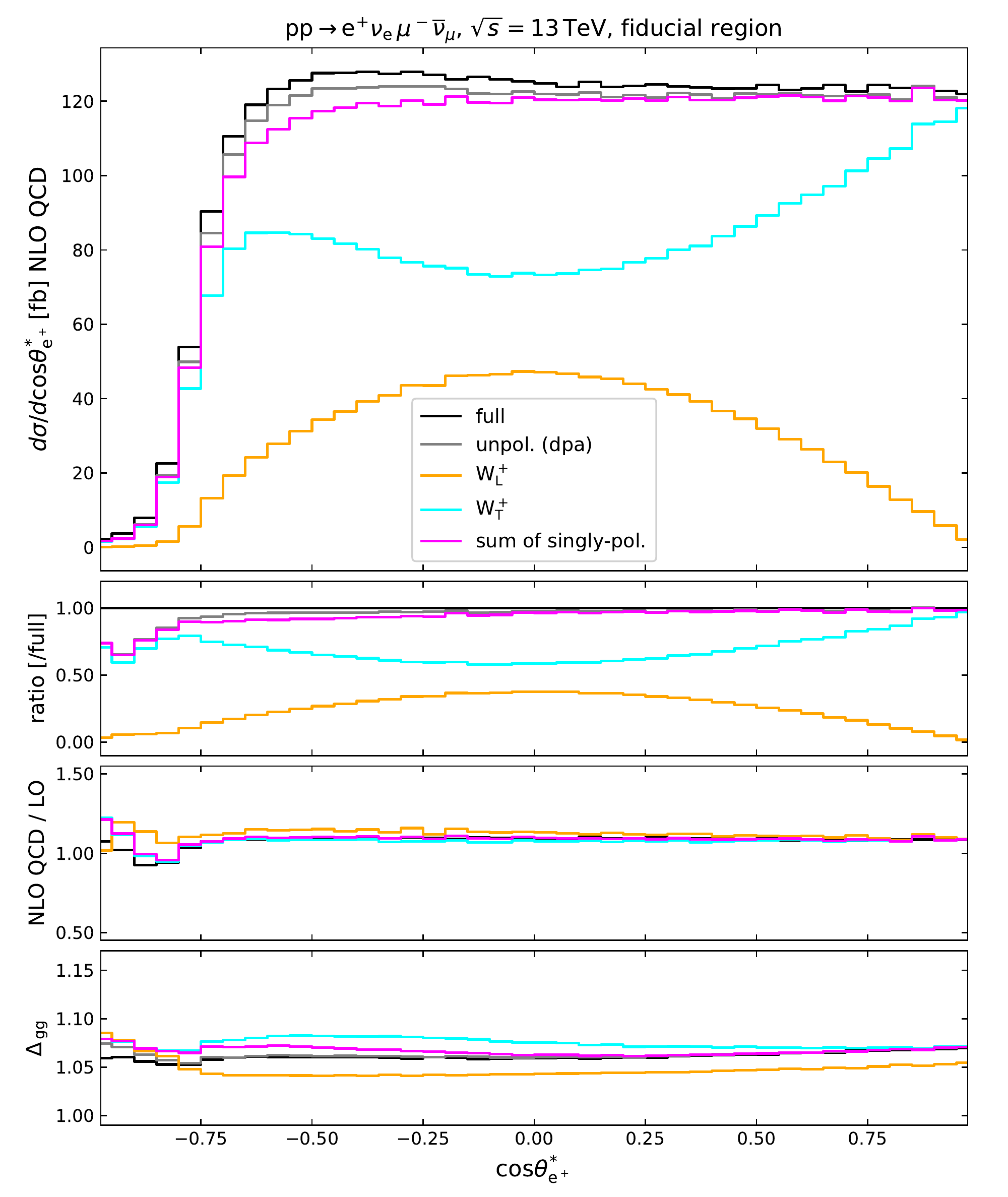}}
  \subfigure[Azimuthal angle of $\Pe^+$ in the $\PW^+$ CM frame.\label{phistar_lep}]{\includegraphics[scale=0.36]{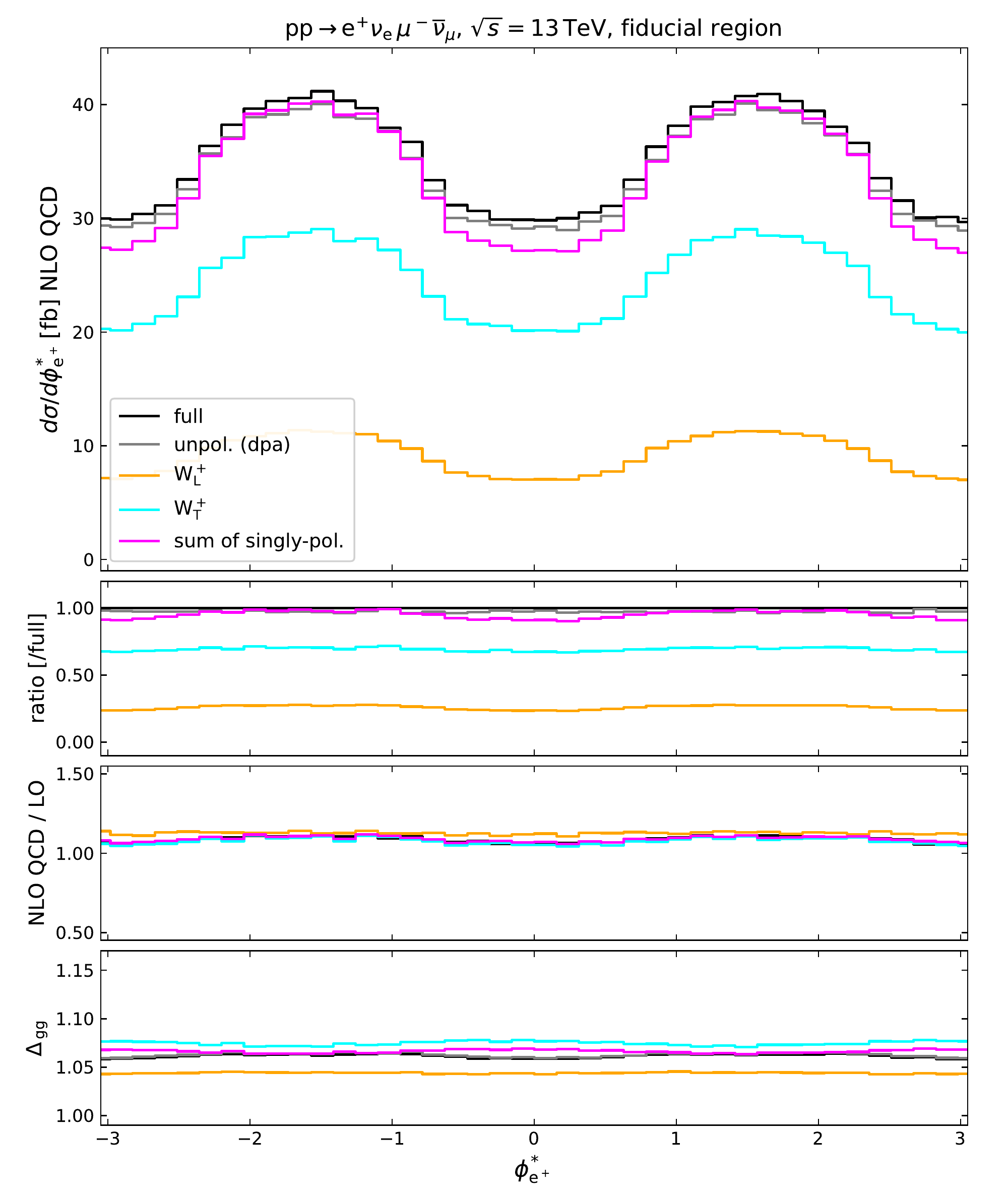}}            
  \caption{Distributions in the positron angular variables
    $\cos\theta^*_\Pe$ and $\phi^*_\Pe$ computed in the $\PW^+$ rest
    frame in the fiducial region. Singly-polarized and unpolarized
    results are shown. Same subplot structure as in
    \fig{stardistribINC1}.}\label{variables2_lep} 
\end{figure}

\begin{figure}
  \centering
  \subfigure[Transverse momentum of the $\PW^+$ boson.\label{ptwp}]{\includegraphics[scale=0.36]{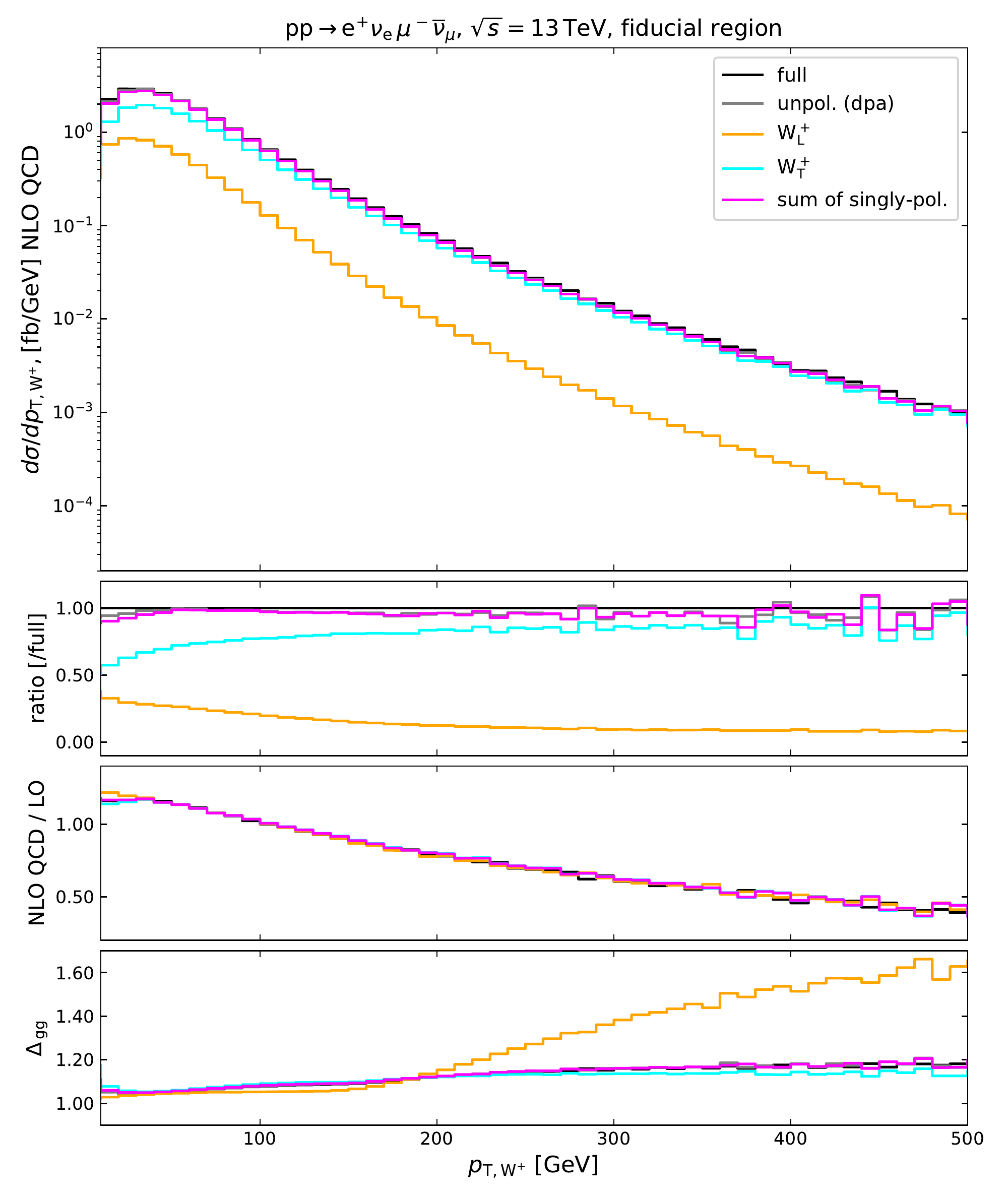}}
  \subfigure[Transverse momentum of the positron.\label{ptep}]{\includegraphics[scale=0.36]{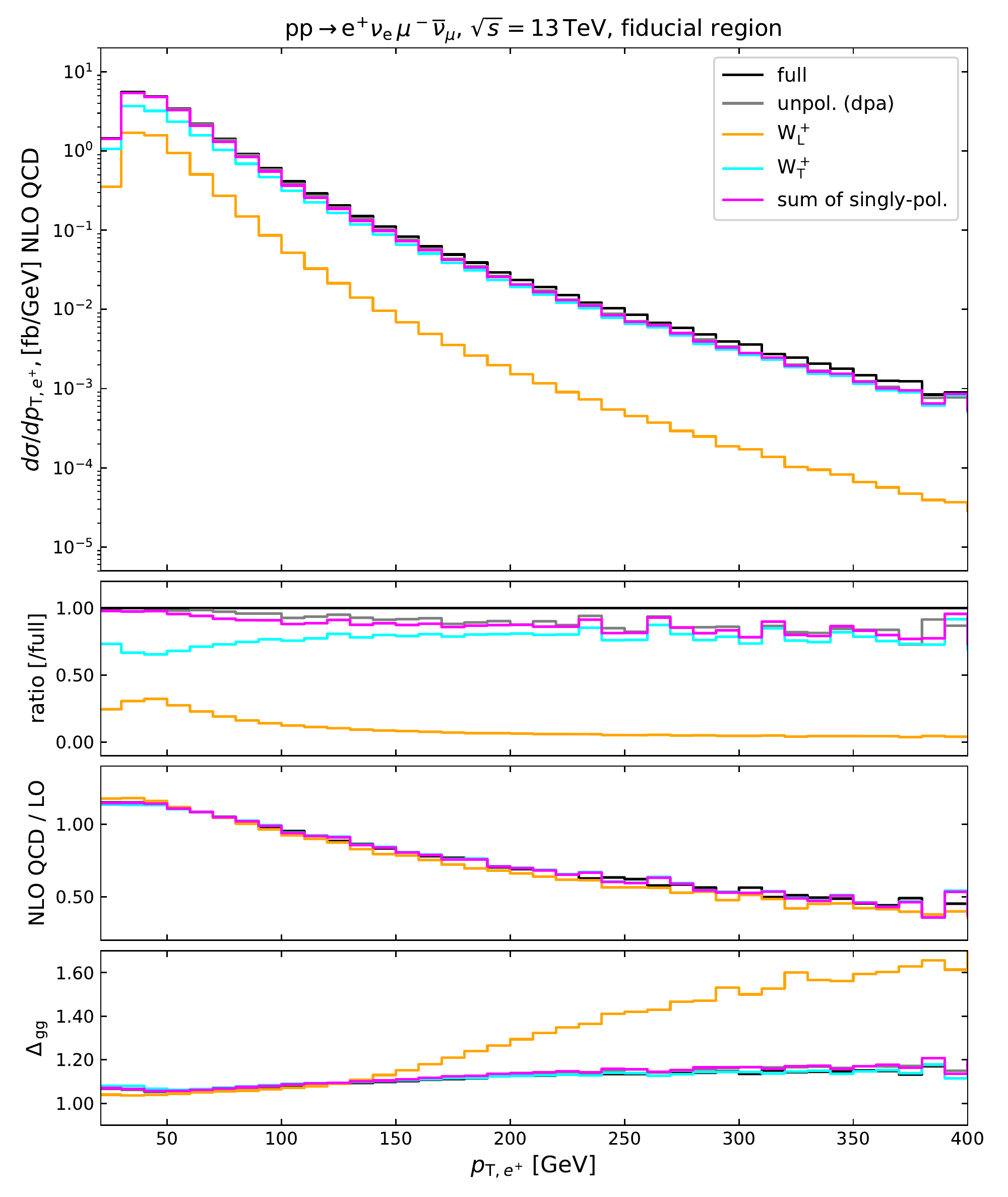}}    
  \caption{Distributions in the transverse momentum of the $\PW^+$
    boson and positron in the fiducial region. Singly-polarized and
    unpolarized results are shown. Same subplot structure as in
    \fig{stardistribINC1}.}\label{variables3_lep} 
\end{figure}

We start by showing in \fig{variables2_lep} how fiducial cuts modify
the angular distributions of leptons in the corresponding $\PW$~rest
frame. We recall that these distributions would be optimal to
discriminate among different polarization modes of decayed weak
bosons, if they could be reconstructed at the LHC
(which is not the case).
The effect of lepton cuts is different for the various polarization
modes, as can be observed comparing \fig{variables2_lep} with
\fig{stardistribINC1}. Since the distribution in the transverse
momentum of the \PW~boson [\fig{ptwp}] peaks near
$p_{\rT,\PW}\approx40\GeV\approx\MW/2$, the transverse-momentum cut on
single leptons suppresses the production of charged leptons which
propagate in the opposite direction of the corresponding decayed boson
($\theta^*_{\Pe} \approx \pi$).  This effect causes the drastic
reduction of polarized cross-sections near $\cos\theta^*_\Pe=-1$,
while other cuts contribute to the suppression at larger
$\cos\theta^*_\Pe$.  Thus, the cuts strongly modify the
transverse-momentum distribution of the positron, in particular, the
component of a left-handed $\PW^+$~boson, which would be maximal for
$\cos\theta^*_\Pe = -1$ in the absence of cuts. The cuts also destroy the
symmetry of the distribution about $\cos\theta^*_\Pe = 0$ for a
longitudinal $\PW$~boson.

An expected consequence of the cuts is the presence of non-vanishing
interferences among polarization modes, which are mostly evident in
the phase-space regions directly affected by the cuts. This is the
case for the negative region of the $\cos\theta^*_\Pe$ distribution
where they account for 5--7\%, as can be extracted by comparing the
violet and gray curves in \fig{costhstar_lep}. Moreover, the effect of
the non-resonant irreducible background is large near
$\cos\theta^*_\Pe = -1$, where a sizeable fraction of the
doubly-resonant contributions is cut away.  For $\cos\theta^*_\Pe >0$
the interferences are less than 2\%, and the DPA is behaving well.

The $K$-factor is roughly equal for the polarized and unpolarized
cases. As already seen at the integrated level, the gluon-induced
partonic channel has a different effect on the spin modes of the
$\PW^+$~boson, enhancing the cross-section for transverse polarization
by $+7.5\%$ but the longitudinal one by $+4\%$.

The $\phi^*_\Pe$ distribution shown in \fig{phistar_lep} features the
same $\cos 2\phi^*_\Pe$ modulation as in the inclusive setup
(\fig{stardistribINC2}) but with a relatively larger amplitude, which
is now present also in the longitudinal component. The agreement
between the unpolarized DPA and the full distribution is very good and
independent of $\phi^*_\Pe$ over the whole range, while the
interferences are a bit enhanced in the minima of the distribution,
where they amount to 8\% of the full result. The NLO QCD corrections
and the \Pg\Pg-channel contribution 
are in line with the integrated results.

We now present transverse-momentum variables. We have already
investigated missing-$\pt$ distributions in the inclusive setup for
polarized \PW~bosons. Because of the bad description of the
unpolarized process by the DPA, this variable is not well suited for
the extraction of polarized signals. 
Therefore, we show no results for this in the fiducial region.

In \fig{variables3_lep} we provide the distributions in the transverse
momentum of the $\PW^+$~boson (not observable) as obtained from Monte
Carlo truth and of the positron (observable).
Since the positron is part of the decay products of the
$\PW^+$~boson, we expect that the polarized results for the two
variables feature similar behaviour. We consider the configuration in
which only the $\PW^+$~boson has a definite polarization state.  First
we estimate the quality of the DPA in the unpolarized case from
\fig{ptwp} and \fig{ptep}.  The transverse-momentum
distribution of the $\PW$~boson is described very well even at large
transverse momentum, while the non-resonant effects become of order
20\% in the tails of the positron transverse-momentum distribution, in
a similar way (but more moderate in size) as for $\ptmiss$.  Up to
this difference which has nothing to do with the polarizations, the
polarized results are similar for $\PW^+$ and $\Pe^+$ both in the
shapes of distributions and in the L/T polarization fractions. The
only difference is due to a mild enhancement of the transverse
component in the soft spectrum of the positron transverse momentum
($20\GeV < {\pte}<50\GeV$). The interferences are very small for the
$\PW^+$ transverse momentum, somewhat larger but always below 5\% for the
positron, which is more directly affected by the lepton cuts.  The
singly-polarized distributions feature differential $K$-factors that
are almost identical to the unpolarized one both in the $\PW^+$ and in
the $\Pe^+$ case. The \Pg\Pg~channel gives roughly the same $10\%$
enhancement to the transverse and to the unpolarized distribution. In
contrast, the longitudinal component is enhanced by more than $50\%$
for $p_{\rT,\PW^+}>400\GeV$ and $\pte>300\GeV$, which means that in
the tails of these distributions the gluonic channel becomes of the
same order of magnitude as the quark-induced one for a longitudinal
$\PW^+$ boson.

Given that off-shell effects are under control and
interferences are moderate, the positron transverse momentum
represents a good observable for polarized signal separation and a
proxy for the $\PW^+$ transverse momentum even for a definite
polarization state.  Similar conclusions can be drawn for the muon and
the $\PW^-$ boson.

\begin{figure}
  \centering
  \subfigure[Rapidity of the $\PW^+$ boson.\label{etawp}]{\includegraphics[scale=0.36]{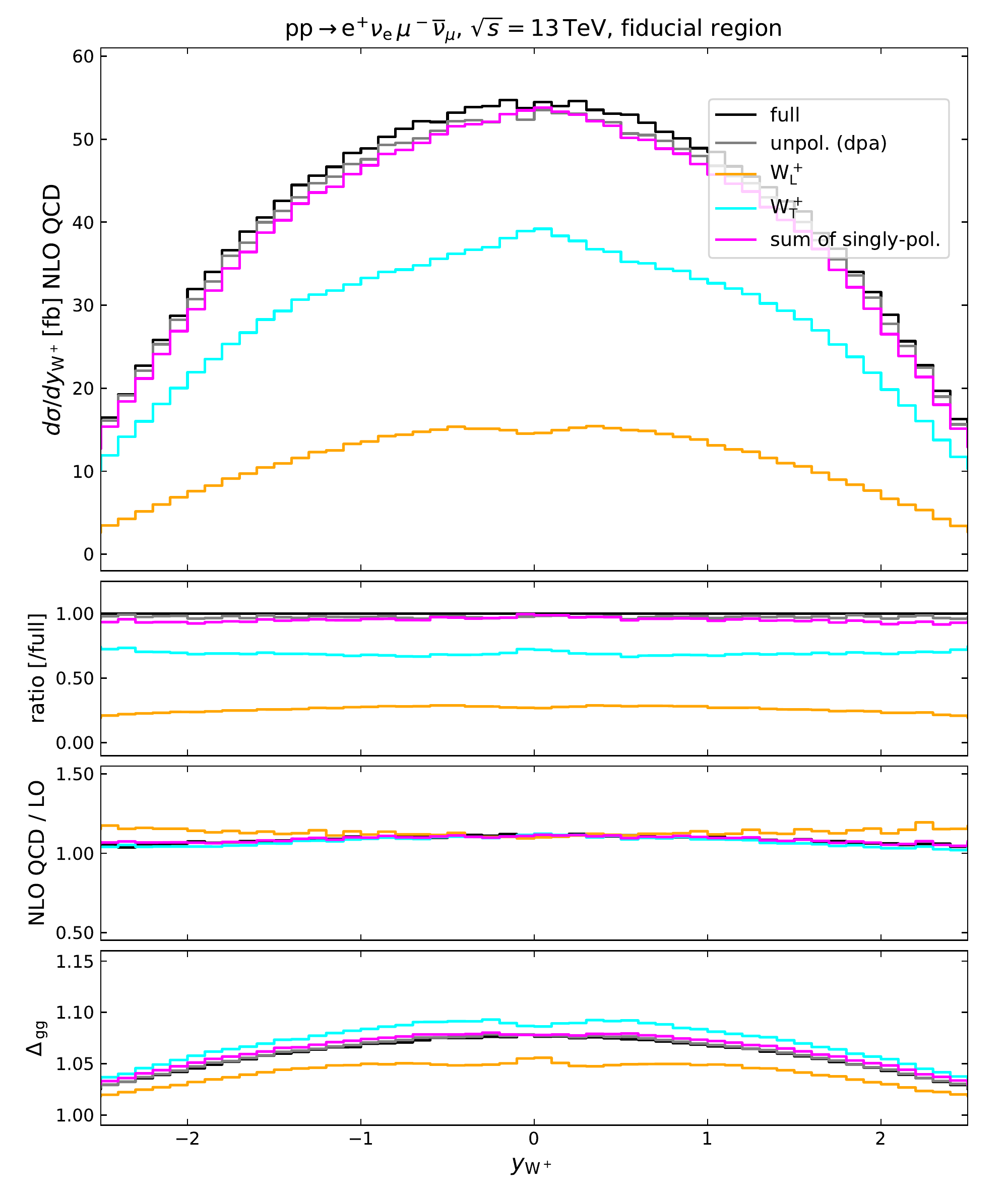}}  
  \subfigure[Rapidity of the positron.\label{etaep}]{\includegraphics[scale=0.36]{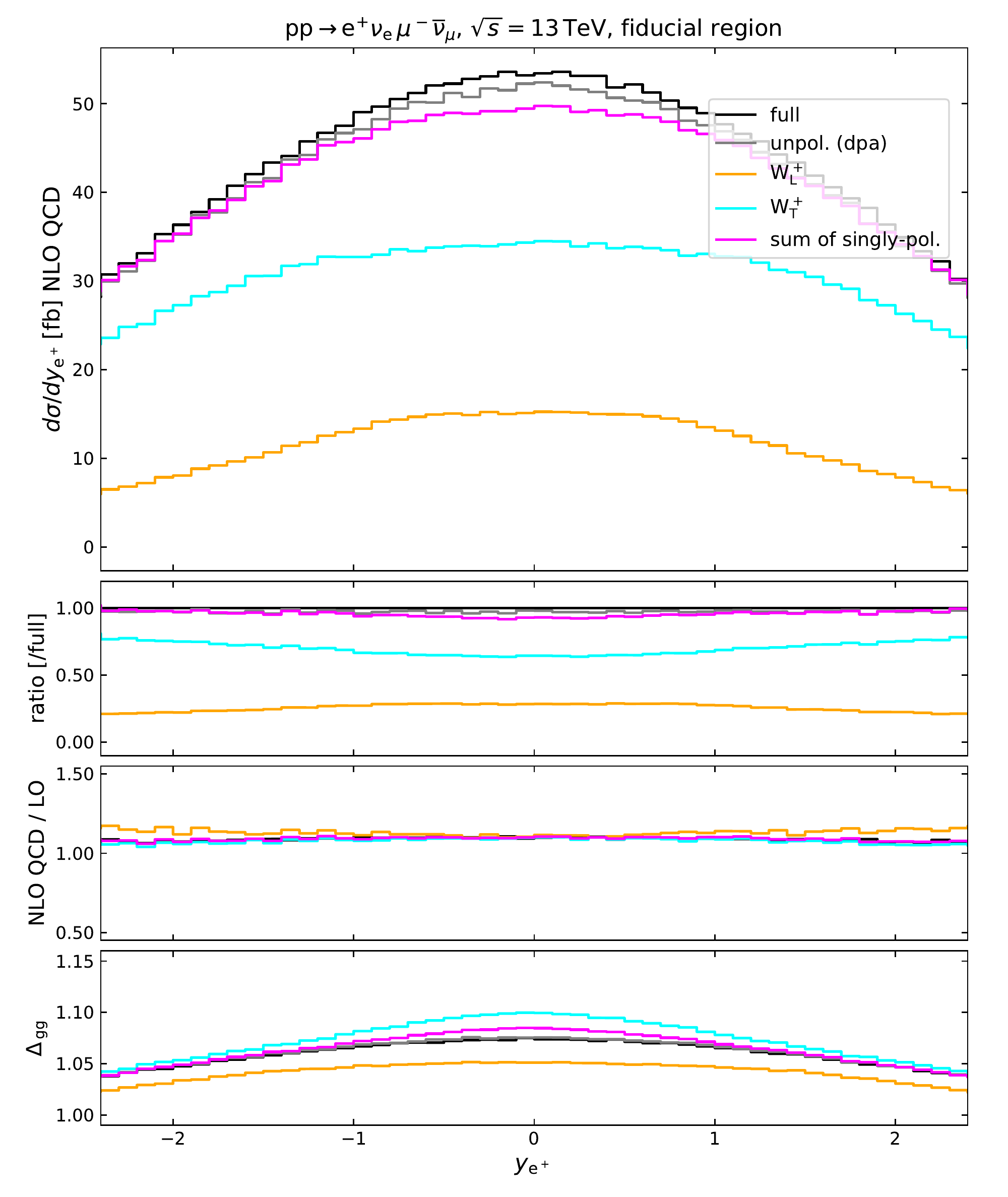}}          
  \caption{Distributions in the rapidity of the $\PW^+$ boson and
    positron in the fiducial region. Singly-polarized and unpolarized
    results are shown. Same subplot structure as in
    \fig{stardistribINC1}.}\label{variables4_lep_single} 
\end{figure}

Another relevant variable linked to a single $\PW$~boson is its
rapidity.  In \figsa{etawp}{etaep} we show the rapidity
distributions for the $\PW^+$~boson and the positron, respectively,
considering a $\PW^+$~boson with definite polarization and an
unpolarized $\PW^-$~boson.
It is evident at first glance that the two variables are directly connected.
The DPA describes the unpolarized full distribution well, and the
interferences are generally small, apart from slightly larger positive
effects in the central region of the positron rapidity (at most 8\%
for $\eta_\Pe^+ = 0$). The $K$-factors for polarized bosons follow the
unpolarized ones, with almost no dependence on the centrality of the
$\PW^+$~boson or the positron.  The gluon-induced process 
contributes the most 
in the central region for both variables.  The
shape of the polarized distributions differs between the
transverse and longitudinal polarizations more in the
distributions of the $\PW^+$~boson. In particular, for $\eta_{\PW^+}=0$ the
transverse distribution is characterized by a small peak, while the
longitudinal one has a local minimum there and peaks
near $\eta_{\PW^+}= \pm 0.4$. This mild effect is reversed in the
gluon-induced process. For the positron rapidity, the transverse and
longitudinal differential cross-sections both feature a maximum in
$\eta_\Pe^+=0$, and most of the differences show up in the distribution
variance, which is slightly smaller in the transverse case.  The
polarization fractions feature very similar behaviours for the two
variables. This gives us further confidence that the positron
kinematics can be used as a proxy for the corresponding $\PW^+$-boson
kinematics at the level of polarized signals.

\begin{figure}
  \centering
  \subfigure[Rapidity of the $\PW^+$ boson.\label{etawpdoub}]{\includegraphics[scale=0.36]{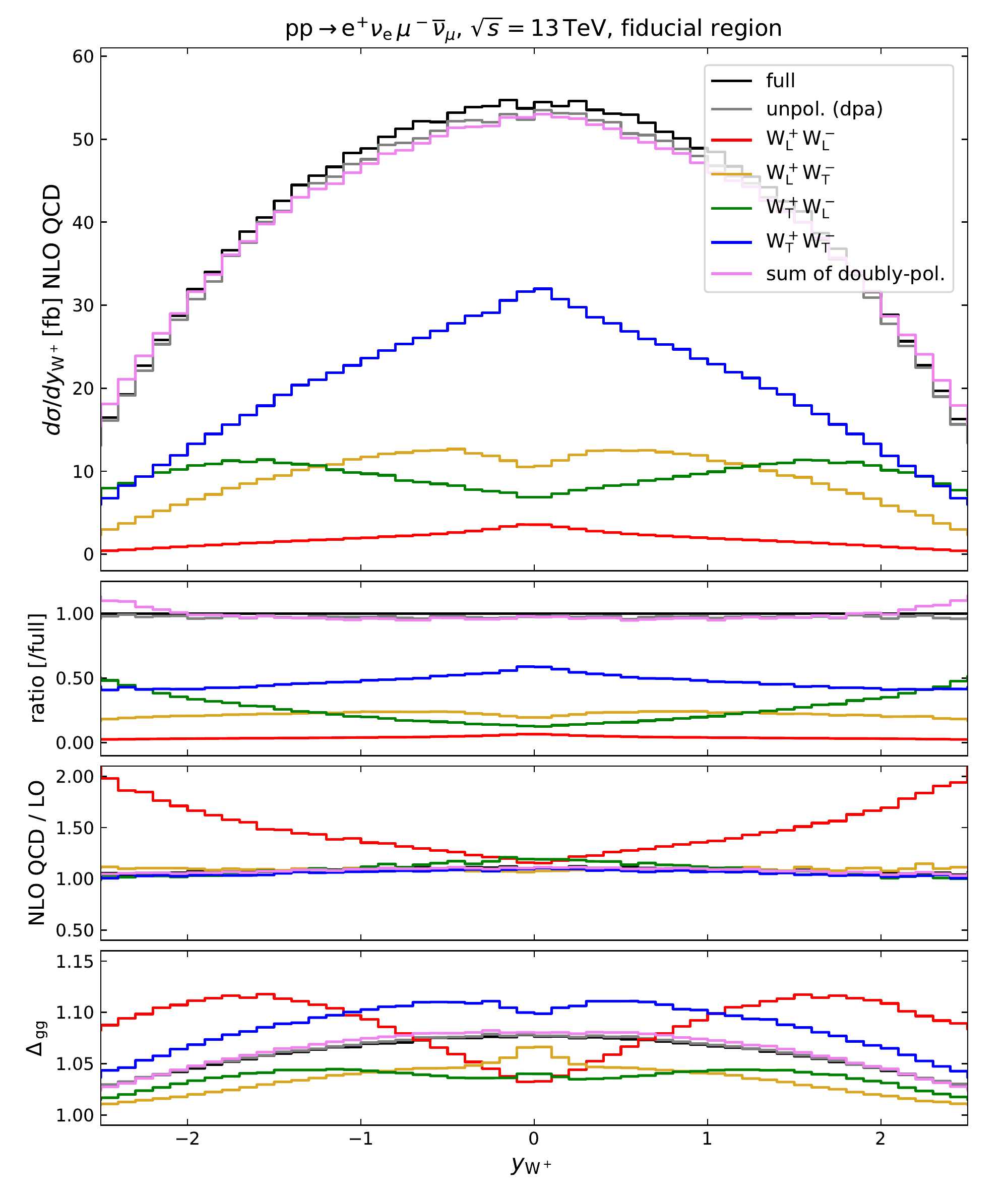}}  
  \subfigure[Rapidity of the positron.\label{etaepdoub}]{\includegraphics[scale=0.36]{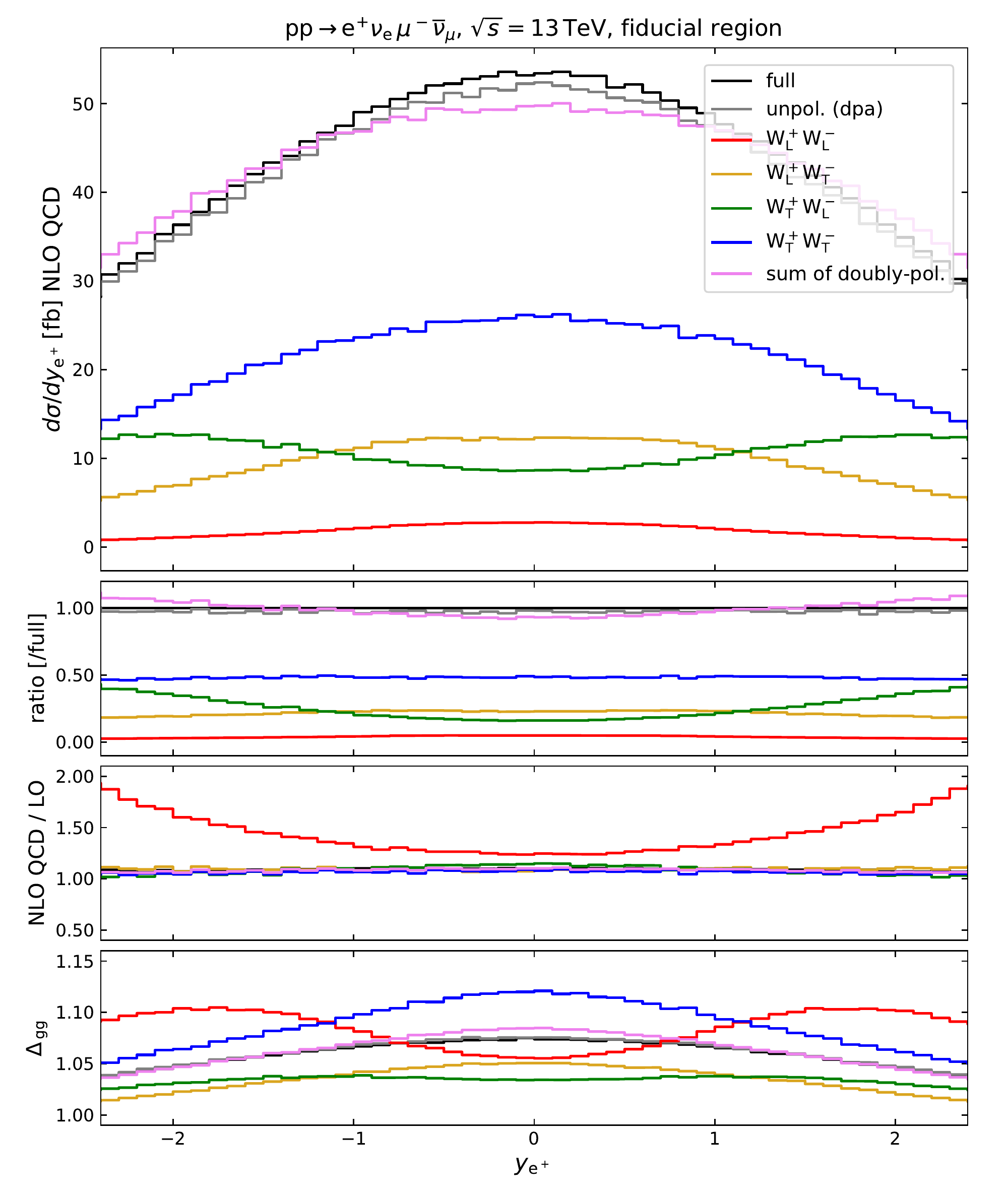}}          
  \caption{Same as in \fig{variables4_lep_single} but for doubly-polarized distributions.}\label{variables4_lep_double}
\end{figure}

In addition to the singly-polarized results in
\fig{variables4_lep_single}, we present in \fig{variables4_lep_double}
the doubly-polarized distributions for the same variables in order to
get information about the correlation between the spin states of the
two bosons.
The interferences are significantly larger than in the
singly-polarized configuration at large rapidities of the $\PW^+$
boson.  Summing over the four combinations of definite polarization
states for both \PW bosons means neglecting the interferences for both
bosons. This results in negative interferences of the order of $10\%$
for $|\eta_{\PW^+}|> 2$.  The interferences are smaller in the
positron distributions, though slightly larger than in the
singly-polarized case.  The doubly-polarized $K$-factors are similar
to the unpolarized one, apart from the LL one.  In
this latter polarization state, the QCD radiative corrections are much
larger at forward or backward rapidities, where they reach 100\%.
However, the LL cross-section is strongly suppressed with respect to
other polarization combinations, as already seen for the fiducial
cross-section.  The \Pg\Pg~channel enhances the
LL distribution, in particular, in the region
$1.5<|\eta|<2$ both for the $\PW^+$ and for the $\Pe^+$. Its
contribution to the TT cross-section peaks at
$|\eta_{\PW^+}|\approx0.5$ and at $\eta_{\Pe^+}=0$. The mixed
combinations receive the smallest enhancement by this partonic
channel.

The most interesting aspect of the doubly-polarized distributions
concerns their shapes, which give much more information than the
singly-polarized ones. As a general statement, the positron rapidity
distributions look like a smoothed version of the $\PW^+$ rapidity
ones.  
The distribution for a transversely polarized $\PW^+$~boson changes
drastically depending on whether the $\PW^-$~boson is longitudinal or
transverse.  
In the former case, the $\PW^+$ rapidity distribution
peaks at $|\eta_{\PW^+}|\approx 1.7$ and has a local minimum for
$\eta_{\PW^+}=0$, while in the latter case, the maximum is at zero
rapidity. Similar comments hold for the $\Pe^+$ rapidity distribution.
Note that for $2.0<|\eta_\Pe^+|< 2.5$ the TL component becomes of the
same order of magnitude as the TT one. These correlation effects could
be helpful in discriminating experimentally between the boson
polarization modes.  The other mixed distribution ($\PW^+$
longitudinal, $\PW^-$ transverse) features two symmetric peaks in
$|\eta_{\PW^+}| \approx 0.4$ and a local minimum at $\eta_{\PW^+}=0$.
However, this does not correspond to an analogous behaviour in the
rapidity distribution of the positron, which is almost flat in the
region $|\eta_{\Pe^+}|< 0.5$.  The LL $\eta_\Pe^+$ distribution has a
maximum at zero~rapidity, which is much less pronounced than the
corresponding maximum of the $\eta_{\PW^+}$ distribution.
Figures \ref{variables4_lep_single} and \ref{variables4_lep_double} show
that extending the investigation to doubly-polarized signals is
definitely needed to completely understand the spin structure in
di-boson production beyond the extraction of the single-boson angular
coefficients.

\begin{figure}
  \centering
  \subfigure[Invariant mass of the $\PW^+$ boson.\label{mww}]{\includegraphics[scale=0.36]{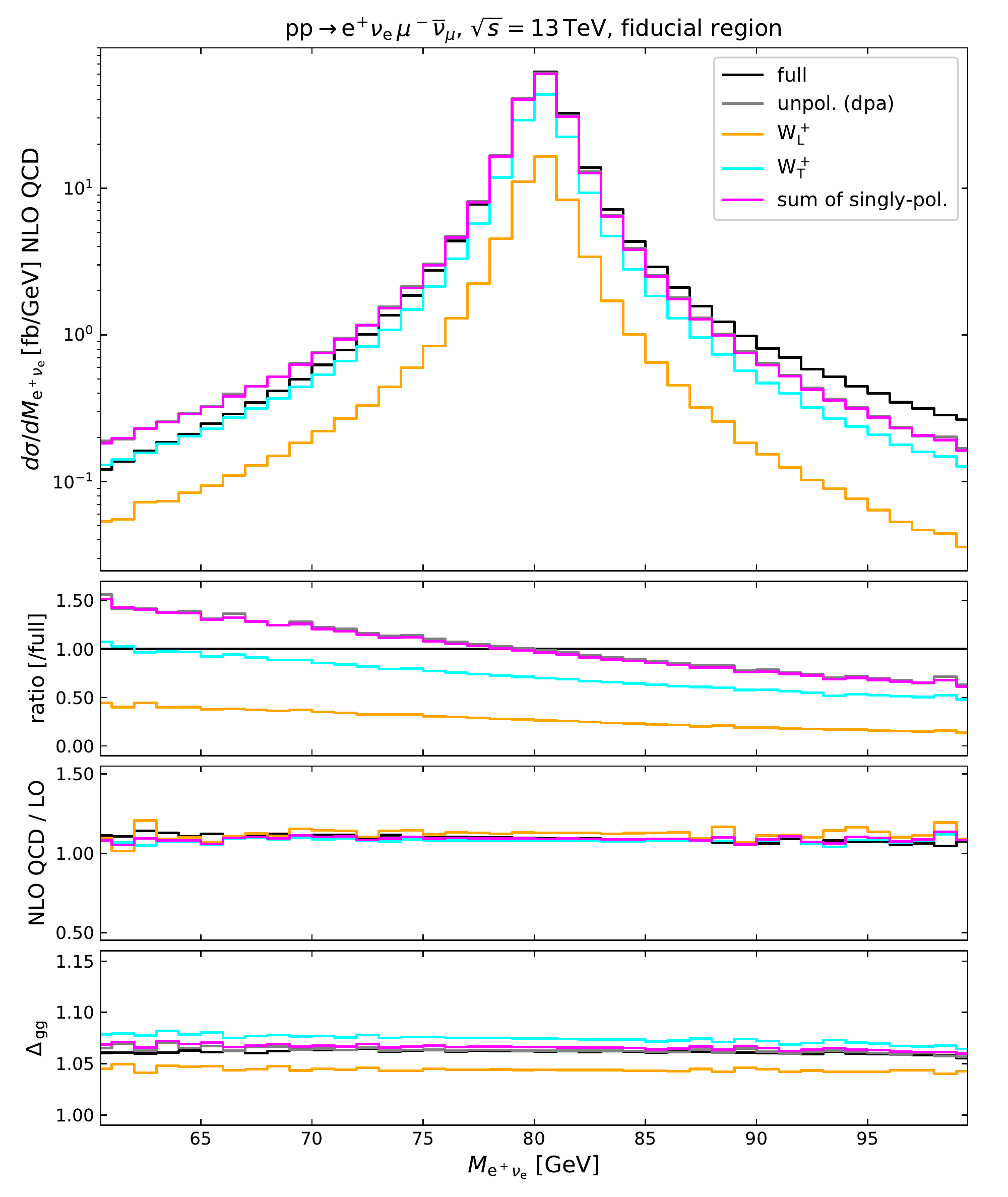}}            
  \subfigure[Invariant mass of the two-charged-lepton system.\label{mlll}]{\includegraphics[scale=0.36]{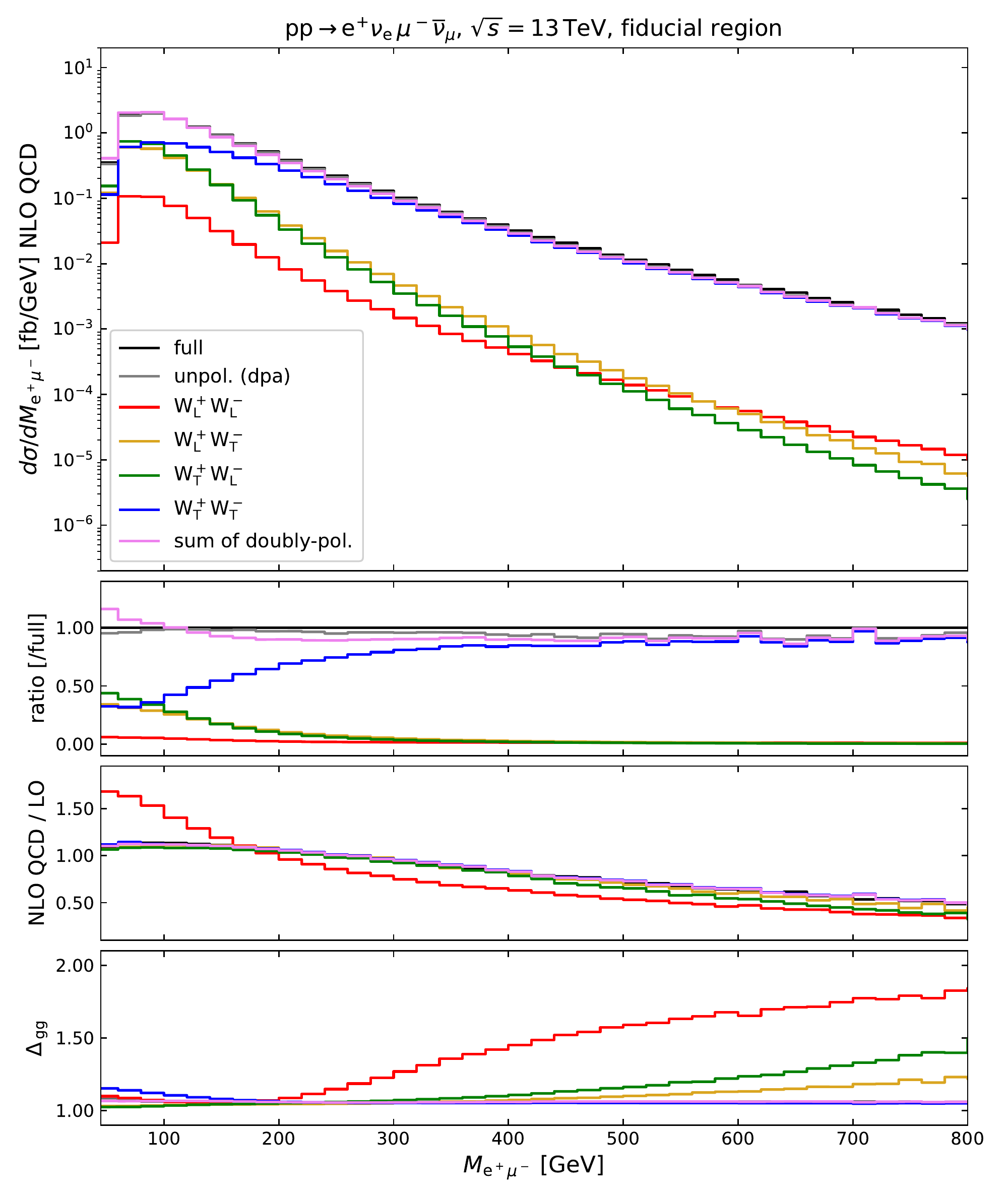}}
  \caption{Distributions in the invariant masses of the $\PW^+$~boson
    (from Monte Carlo truth) and of the two charged leptons in the fiducial
    region. Singly-polarized results are shown in \fig{mww},
    doubly-polarized ones in \fig{mlll}.  Same subplot structure as in
    \fig{stardistribINC1}.}\label{variables5_lep}
\end{figure}

We present in \fig{mww} the distribution in the invariant $\PW^+$ mass
reconstructed from Monte Carlo truth by summing the positron and
electron-neutrino momenta.
Though not observable at the LHC, this variable provides interesting
information about the DPA description of the full kinematics. The full
distribution, which includes all resonant and non-resonant diagrams,
is not symmetric about the $\PW$~pole mass, but favours invariant-mass
values larger than $\Mw$. The employed DPA technique projects the
kinematics of the amplitude numerator on the mass shell, preserving
off-shell kinematics in the (symmetric) Breit--Wigner modulation in
$\PW$-boson propagators. This renders the invariant-mass distributions
more symmetric about the pole mass, as can be seen in \fig{mww}. The
discrepancy between the approximated and full results is in fact
positive for $M_{\Pe^+\nu_\Pe}<\Mw$ and negative otherwise.  It
reaches $\pm 50\%$ for $M_{\Pe^+\nu_\Pe} = \Mw \mp 20\GeV$.  The
polarized distributions have more or less the same shape, and their
sum reproduces almost perfectly the unpolarized DPA results, in spite of
the application of lepton cuts.  All these considerations hold with
no modification at LO for both the $\qqb$ and the gluon-induced
partonic processes as well as at NLO QCD for the quark-induced
process.

The invariant mass of the system formed by the two charged leptons is
definitely observable at the LHC. The corresponding doubly-polarized
distributions are shown in \fig{mlll}.  The DPA reproduces the full
result reasonably well, in particular in the region
$M_{\Pe^+\mu^-}<400\GeV$ where the discrepancies between the two
unpolarized predictions are below 10\%. In the same region, the
polarization interferences are of order of 10\%, negative fo
$M_{\Pe^+\mu^-}<100\GeV$ and positive between 100 and $300\GeV$.  In
the tails of the distributions, the non-resonant effects dominate the
discrepancy with respect to the full result (15\%), while the
interferences are negligible.  The same effects can be found even in
the inclusive setup, as well as in the study of singly-polarized
distributions.  It seems likely to be related to the strong
correlation between the bosons in $\PW$-pair production, as well as to
the binning of the distributions which introduces implicit cuts on the
variable itself (as for other leptonic kinematic variables).
For $M_{\Pe^+\mu^-}> 500\GeV$ the LL component is
larger than the mixed ones. However, all of the three combinations
involving at least one longitudinal boson are strongly suppressed
in the high-mass region (two orders of magnitude smaller than the
doubly-transverse one).  In the soft region of the spectrum (where
interferences are sizeable) the TL/LT contributions are even larger
than the TT one.  Given the limited
experimental statistics, the soft part of the spectrum is the only one
which is accessible and worth investigating. Furthermore,
given the Higgs-background cut $M_{\Pe^+\mu^-}>55\GeV$ imposed on this
variable, it would be
interesting to study the effect of varying such a cut on the polarized
distributions.  The $K$-factors are below one for invariant masses
larger than 200 GeV, and the LL one decreases faster
than the others. On the contrary, in the soft region, $K$-factors are
above one and the LL one is much higher than the others.
The \Pg\Pg~channel enhances mostly the configurations with at least
one longitudinal boson for $M_{\Pe^+\mu^-}> 200\GeV$. In the tails of
the distributions this contribution becomes of the same order of
magnitude as the quark-induced one for the LL signal.

\begin{figure}
  \centering
  \subfigure[Azimuthal separation between $\Pe^+$ and $\mu^-$.\label{azimuthemu_lep}]{\includegraphics[scale=0.36]{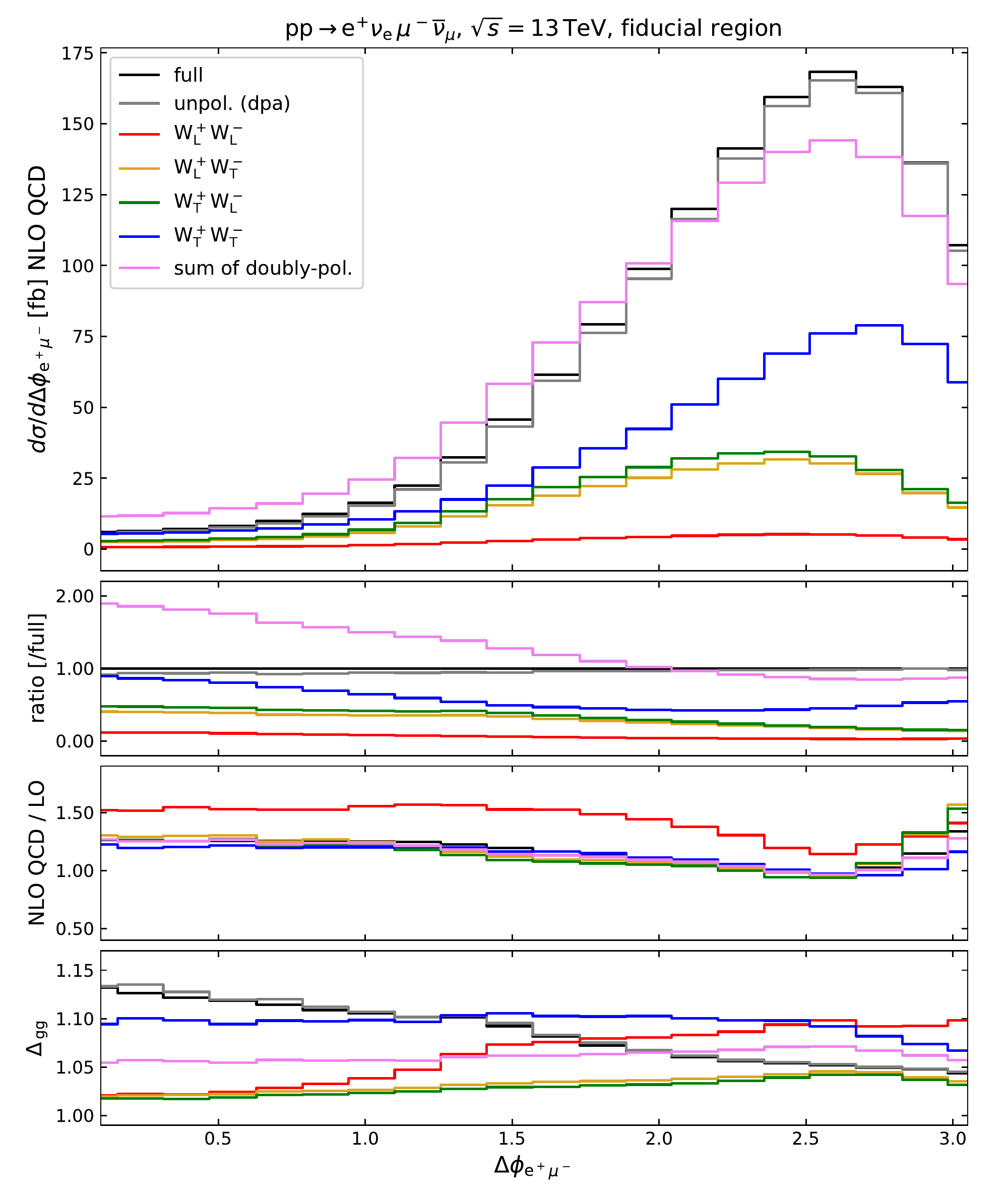}}
  \subfigure[Cosine of angle between $\Pe^+$ and $\mu^-$.\label{costhemu_lep}]{\includegraphics[scale=0.36]{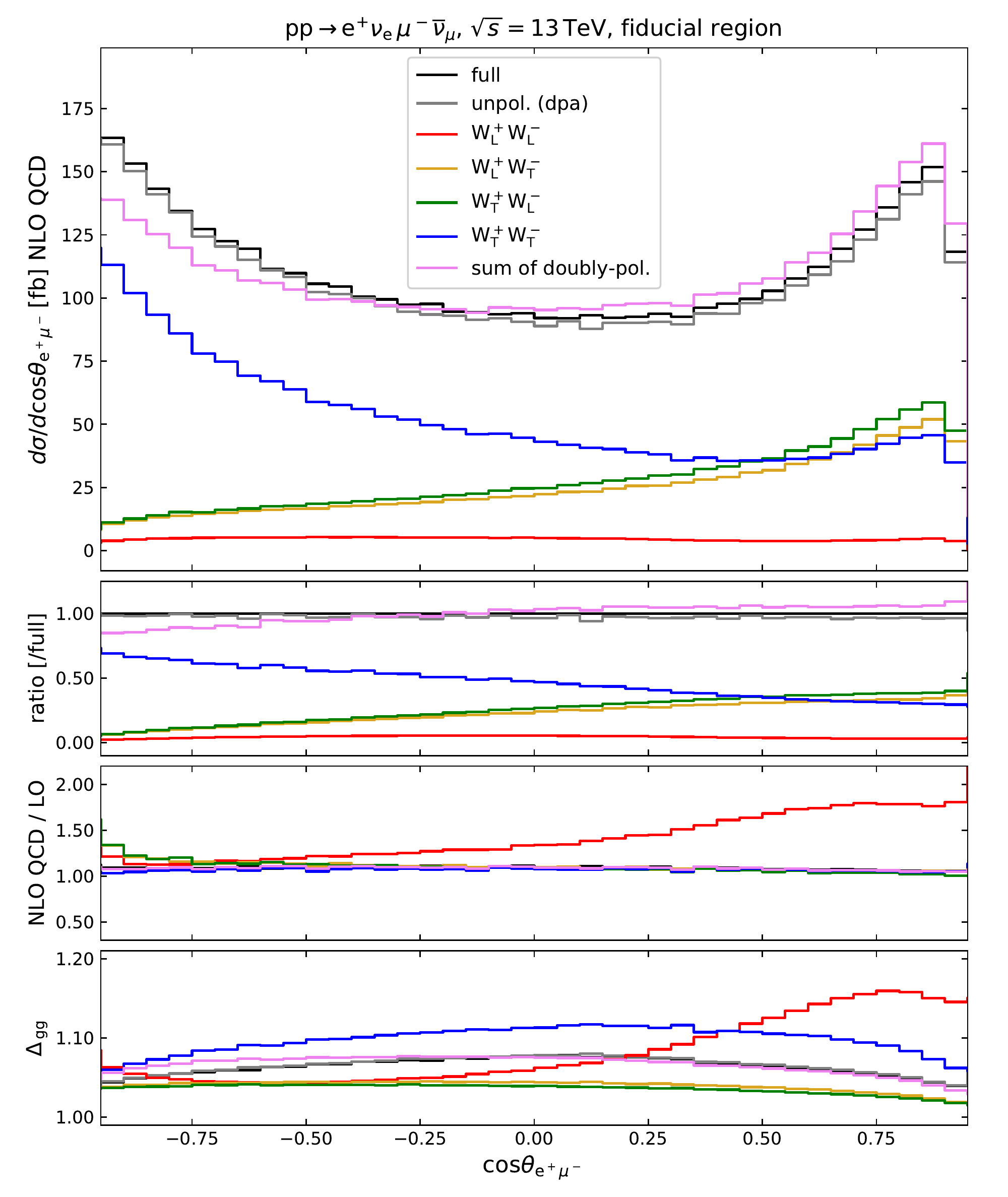}}            
  \caption{Distributions in the azimuthal separation and in the cosine
    of the angle between the two charged leptons
    in the fiducial region. Doubly-polarized results are shown.  Same
    subplot structure as in
    \fig{stardistribINC1}.}\label{variables1_lep}
\end{figure}

In \fig{variables1_lep} we present doubly-polarized distributions for
two different angular variables between the two charged leptons. 
In the azimuthal separation [\fig{azimuthemu_lep}], which we already
investigated in the inclusive setup in \fig{azimuthll}, the
distributions in the presence of lepton cuts feature a peak near
$\Delta\phi_{\Pe^+\mu^-}\approx 2.6$ (it is $\pi$ in the inclusive
setup).  The interference effects discussed in \sect{sub:inc} are
strongly enhanced and reach almost 100\% for $\Delta\phi_{\Pe^+\mu^-}$
close to zero.  Apart from the modification of the shapes, the lepton
cuts do not change the conclusions that we have drawn in the
inclusive setup.  A possible improvement in the description of this
variable could only be given by a cut. Since the Higgs signal is
enhanced for $\Delta\phi_{\Pe^+\mu^-}<1.8$ \cite{ATLAS:2014aga},
selecting the complement of the spectrum helps extracting more
precisely the di-boson signal.  Note that imposing
$\Delta\phi_{\Pe^+\mu^-}>1.8$ would also mean excluding the region
which is dominated by large negative interferences.

In \fig{costhemu_lep} the distribution in $\cos\theta_{\Pe^+\mu^-}$
(computed in the laboratory frame) is presented. At the unpolarized
level, the two leptons tend to be produced in a collinear
configuration, if lepton cuts are absent. This is in agreement with
the fact that the positron is preferably produced in the opposite
direction of the $\PW^+$~boson, while the muon is produced mostly in
the same direction of the $\PW^-$~boson, as can be also deduced from
$\cos\theta^*_\ell$ distributions. However, the application of
selection cuts on the leptons impedes the collinear configuration
($\cos\theta_{\Pe^+\mu^-}=1$) and the resulting situation is the
following: the TT contribution has its maximum at
$\cos\theta_{\Pe^+\mu^-} = -1$, while the mixed contributions peak at
values slightly smaller than $\cos\theta_{\Pe^+\mu^-} = 1$. Note that the
TL and LT contributions have the same shape but differ in the overall
normalization according to the total cross-sections shown in
\tab{table:sigmaleptNLOvetoed}.  The LL distribution vanishes for
collinear leptons and is almost flat in the rest of the spectrum up to
a very mild tendency to prefer negative values of
$\cos\theta_{\Pe^+\mu^-}$.  The DPA reproduces very well the full
computation in this distribution. The interferences are moderate and
feature a change of sign at $\cos\theta_{\Pe^+\mu^-}=0$. For positive
values of the variable they are negative and account at
most for $10\%$, while for negative values they are positive and
amount to $15$--$20\%$ in the back-to-back configuration.  The QCD
corrections enhance the mixed and LL contributions by
roughly $20\%$ for $\cos\theta_{\Pe^+\mu^-}\approx -1$.  Furthermore, the
LL distribution benefits from the radiative
corrections in the positive side of the spectrum, where the
corrections are again of order $50\%$. It receives a similar enhancement
of up to $15\%$ also from the combination with the gluon-induced
channel.  This partonic process enhances the TT
signal by more than 10\% in the central part of the distribution.
The good DPA description of the unpolarized cross-section, the
presence of interferences that are sizeable but can be taken into
account in the SM, and the clear differences in the shapes of
polarized distributions make this angular observable a good candidate
for the discrimination of polarized signals at the LHC.

\section{Conclusion}\label{sub:con}
In this paper we have studied $\PW$-pair production at the LHC with
one or both bosons in definite polarization states.  We have included
NLO QCD corrections to the leading $\qqb$ partonic process
as well as the loop-induced gluon-initiated contribution at LO.

The polarized signals are defined at the amplitude level and rely on
the double-pole approximation. The doubly-resonant contributions are
separated in a gauge-invariant way at LO and at NLO QCD, including the
(integrated and unintegrated) subtraction counterterms and the real
corrections. Amplitudes for polarized vector bosons are defined based
on the gauge-invariant doubly-resonant contributions in the
double-pole approximation.  This strategy is radically different from
other methods that have been used in the literature to define
polarized cross-sections for unstable particles.  In particular, this
technique allows one to define polarized cross-sections while
retaining some off-shell effects and all spin correlations.  We have
evaluated the quality of thus-defined polarized signals both in terms
of the missing off-shell effects and in terms of the interferences
among polarization states.
The results are not limited to the polarization of a single boson but
target a more complete description of the spin structure of the
process by means of the doubly-polarized signals, which give access to
the correlation between the polarization modes of the two bosons.

We have presented total and differential cross-sections both in an
inclusive setup and in a fiducial region that mimics the one of the
most recent ATLAS measurement in $\PW^+\PW^-$ production.
The inclusive setup serves as a validation framework: the comparison
with results extracted from unpolarized distributions via projection 
on polarized angular distributions gives very good agreement for both
singly- and doubly-polarized signals.  As a by-product, we have found
that already in the inclusive setup some observables are subject to
large interferences and non-resonant background effects, and thus  not
well suited to extract the weak-boson polarizations.
In the fiducial region, we have investigated the effect of a realistic
set of cuts on distributions for polarized and unpolarized bosons
with the aim to identify the observables that are particularly sensitive to
the polarization of decayed bosons, and more in general to understand
how the polarization selection modifies the distributions with respect
to the unpolarized case.

The polarization fractions are very stable against scale variations,
and the related theoretical error is usually at the sub-percent level
both at LO and NLO QCD.  The NLO-QCD $K$-factors for singly-polarized
processes are very close to the ones of the full computation. The same
holds for doubly-polarized cross-sections that feature at least one
transverse $\PW$~boson. The distributions for purely longitudinal
$\PW$~bosons receive very large $K$-factors despite the application of
a jet veto.

From the combination of the quark- and gluon-induced contributions, we
verified that the spin structure of the initial state influences
the final-state polarization modes considerably owing to the limited
number of final-state particles.

This represents a first realistic study of vector-boson polarizations
in $\PW^+\PW^-$ hadronic production, which will hopefully help
addressing future experimental analyses that target the extraction of
polarized signals from LHC data.

\section*{Acknowledgements}
We are grateful to Jean-Nicolas Lang for supporting \recola
and to Timo Schmidt and Mathieu Pellen for maintaining \mocanlo.
GP thanks Alessandro Ballestrero and Ezio Maina for useful
discussions.  The authors acknowledge financial support by the German
Federal Ministry for Education and Research (BMBF) under contract
no.~05H18WWCA1.

\bibliographystyle{JHEPmod}
\bibliography{polvv}

\end{document}